\documentclass[12pt,preprint]{aastex}
\usepackage{natbib}

\def\epsir{\epsilon_{\rm IR}}
\def\spit{{\it Spitzer}}
\def\rg{{r_p}}

\def\rgmax{{r_{\rm max}}}
\def\70um{70~\micron}
\def\160um{160~\micron}
\def\24um{24~\micron}
\def\um{\micron}
\def\ld{L_{\rm dust}/L_{\star}}

\def\Tstar{T_{\star}}

\def\MEarth{M_\oplus}

\def\gapp{\lower 3pt\hbox{${\buildrel > \over \sim}$}\ }
\def\lapp{\lower 3pt\hbox{${\buildrel < \over \sim}$}\ }
\def\proptosim{\lower 3pt\hbox{${\buildrel \propto \over \sim}$}\ }

\def\arcsec{$^{\prime\prime}$}

\begin{document}

\title{Frequency of Debris Disks around Solar-Type Stars: \\
First Results from a \spit/MIPS Survey}

\author{
G. Bryden$^{1}$, C. A. Beichman$^{2}$, D. E. Trilling$^{3}$, 
G. H. Rieke$^{3}$, E. K. Holmes$^{1,4}$, S. M. Lawler$^{1}$,
K.~R. Stapelfeldt$^{1}$, M. W. Werner$^{1}$, T.~N. Gautier$^{1}$,  
M. Blaylock$^{3}$, K. D. Gordon$^{3}$, 
J. A. Stansberry$^{3}$, \& K. Y. L. Su$^{3}$
}
\affil{1) Jet Propulsion Lab, 4800 Oak Grove Dr, Pasadena, CA 91109} 
\affil{2) Michelson Science Center, California Institute of Technology, 
  Pasadena, CA 91125}  
\affil{3) Steward Observatory, University of Arizona, 
  933 North Cherry Ave,  Tucson, AZ 85721} 
\affil{4) Deceased, 2004 March 23.}

\shorttitle{Debris Disks around Solar-Type Stars}
\shortauthors{Bryden et al.}

\begin{abstract}

We have searched for
infrared excesses around a well defined sample of 69 FGK main-sequence
field stars.  These stars were selected without regard to their age,
metallicity, or any previous detection of IR excess; they have a
median age of $\sim$4 Gyr.  We have detected \70um excesses around 7
stars at the 3-$\sigma$ confidence level.
This extra emission is produced by cool material ($<100$~K) located
beyond 10~AU, well outside the ``habitable zones'' of these systems and
consistent with the presence of Kuiper Belt analogs with $\sim$100 times
more emitting surface area than in our own planetary system.  
Only one star, HD~69830, shows excess emission at \24um, corresponding
to dust with temperatures $\gapp$300~K located inside of 1~AU. 
While debris disks with $\ld \geq 10^{-3}$ are rare around old FGK
stars, we find that the disk frequency increases from 2$\,\pm\,$2\% for
$\ld \geq 10^{-4}$ to 12$\,\pm\,$5\% for $\ld \geq 10^{-5}$. 
This trend in the disk luminosity distribution is consistent with 
the estimated dust in our solar system being 
within an order of magnitude, greater or less,
than the typical level around similar nearby stars. 

\end{abstract}


\section{Introduction}

Based on the low level of infrared emission from dust in the solar
system, the discovery with IRAS of infrared 
emission from debris disks around other main sequence
stars was very unexpected \citep{aumann84}. The dust temperatures in the IRAS-detected extrasolar
debris disks (50-150 K) are similar to those in the solar system,
indicating that the material is at roughly similar distances from the
stars, 1 to 100 AU. However, the strength of the emission is much higher.
Observed dust luminosities range from $\ld \simeq 10^{-5}$ to greater
than $10^{-3}$. In comparison, our solar system
has $\ld \simeq 10^{-7}$ to $10^{-6}$ in the
Kuiper Belt, estimated primarily from
extrapolations of the number of large
bodies \citep{Stern96}, and $10^{-8}$ to $10^{-7}$ for the asteroid
belt, determined from a combination
of observation and modeling \citep{Dermott02}.
Because radiation pressure and Poynting-Robertson drag remove dust
from all these systems on time scales much shorter than the stellar ages,
the dust must have been recently produced.
In the solar system, for example, dust is continually generated by
collisions between larger bodies in the asteroid and Kuiper belts, as
well as from outgassing comets. 

The IRAS observations were primarily sensitive to material around A
and F stars, which are hot enough to warm debris effectively. 
Because IRAS was not in general sensitive to disks
as faint as $\ld \sim 10^{-5}$, most detections were of brighter
debris disks, particularly for the cooler, roughly solar-type stars.
For disk luminosities as low as $\ld \sim 10^{-5}$, the only 
solar-type IRAS detection was $\tau$ Ceti, a G8 star located just 
3.6 pc away.  A general statistical analysis of IRAS data, taking into
account the selection biases, could only constrain
the fraction of main-sequence stars with IR excess to be between 3 and
23\%, at a 95\% confidence level \citep{plets99}.

Most of the initial debris disk discoveries were for stars much
younger than the Sun, suggesting that the lower amount of dust in the
solar system could be explained by a declining trend in
dust luminosity over time \citep{aumann84}.
Observing over a range of spectral types, ISO found such a general decline
\citep{spangler01}, but with
the possibility of finding modest excesses at almost any age
\citep{decin00,decin03,Habing01}.
\spit\ observations of A stars confirm an overall decline in the
average amount of \24um excess emission on a $\sim$150 Myr time scale
\citep{Rieke05}.  On top of this general trend, Rieke et al.\
also find large variations of the excess within each age group,
probably due in part to sporadic replenishment of dust clouds by
individual collisions between large, solid bodies, but also likely a
reflection of a range in mass and extent for the initial planetesimal disk.
The detection of strong IR excesses around A stars $\sim$500 Myr old
\citep{Rieke05}, well beyond the initial decline, suggests that
sporadic collisions around stars even several billion years old might
produce significant amounts of dust.

To place the solar system in context, and also to understand debris disk
evolution beyond the $\sim$ 1 Gyr lifetimes of A and early F stars, requires
understanding the characteristics of debris systems around solar-type stars.
Observations with ISO have helped in this regard. 
\citet{decin00} identified strong 60 \um\ IR excess around 3 out of
30 G-type stars, for a detection rate of 10$\,\pm\,$6\%.
Two of these three detections were previously identified by
IRAS.\footnote{\citet{decin00} also identified two
additional stars with potential IR excess, but noted that depending on
the method of data reduction they might not be real detections.
We find with \spit\ that at least one of the two, HD 22484, is
indeed spurious.}
Based on a more general ISO survey and IRAS data,
\citet{Habing01} compiled a larger sample for
determining the fraction of solar-type stars with IR excess.
Among 63 F5-K5 stars, they identify 7 stars
with significant IR excess giving a detection rate of 11$\,\pm\,$4\%.
All their detections have relatively high 60 \um\ fluxes ($> 100$ mJy).
Despite ISO's noise level of $\sim$30 mJy, by restricting their sample
to the closest stars Habing et al.\ are generally sensitive down to
$\ld$ of several times $10^{-5}$.

IRAS and ISO observations provide important limits on the frequency of
FGK stars with debris disks, but because of limitations in sensitivity 
they can probe only the brightest, closest systems and cannot achieve
adequate detection rates to establish many results on a sound statistical basis. 
The Multiband Imaging Photometer on \spit\ \citep[MIPS;][]{Rieke04}
provides unprecedented sensitivity at far-IR wavelengths  
($\sim$2 mJy at \70um; see \S\ref{noisesec}) and 
is an ideal instrument to extend this work.
It is now possible to measure a large enough sample of solar-type
stars down to photospheric levels to constrain the overall
distribution of debris disks.
\spit/MIPS allows the search for disks around FGK stars to be
extended to greater distances and more tenuous
disks than was previously possible.
                                                                            
The FGK Survey is a \spit\ GTO program designed to search for excesses
around 150 nearby, F5-K5 main-sequence field stars, sampling
wavelengths from 8 to 40 \um\ with IRS and 24 and \70um with MIPS.
This survey is motivated by two overlapping scientific goals: 1) to
investigate the distribution of IR excesses around an unbiased sample
of solar-type stars and 2) to relate observations of debris disks to the
presence of planets in the same system.
Preliminary results for the planet component of our GTO program
are discussed in a separate paper \citep{Beichman05planets};
here we focus on the more general survey of nearby, solar-type stars.
The IRS survey results are presented in \citet{Beichman05IRS},
while the first results of the MIPS 24 and \70um survey are presented below.
A large sample of solar-type stars has also been observed as a \spit\
Legacy program \citep{meyer,kim05}.  That program primarily targets more distant
stars and hence only detects somewhat more luminous excesses, but
provides adequate numbers for robust statistics on such systems.

In \S\ref{sample} we describe our sample selection
based on predicted IR fluxes (\S\ref{kuruczsec}).
We present our MIPS observations in \S\ref{observations}, concentrating
on the sources of background noise and
a thorough error analysis to determine whether the measured excesses are
statistically significant (\S\ref{noisesec}).
In \S\ref{ldsec} we discuss how our MIPS observations constrain the
dust properties in each system.
\S\ref{parameters} shows our 
attempts to find,
for systems with IR excess,
correlations with
system parameters such as stellar metallicity and age.
Finally, based on our preliminary data, we calculate the distribution of
debris disks around solar-type stars and place
the solar system in this context
(\S\ref{frequency}).

\section{Stellar Sample}\label{sample}

The FGK program consists of two overlapping sets of stars: 
those which meet strict selection criteria for an unbiased sample
and those which are known to harbor planets.
In both cases, only stars with spectral type similar to the Sun are
considered. 
Observations of FGK planet-bearing stars have already been presented
in \citet{Beichman05planets}; here we concentrate on the larger, unbiased
sample of nearby solar-type stars.

Among stars with spectral type F5 to K5 and luminosity class IV or V,
our targets are chosen mainly based on the expected signal-to-noise
ratio (S/N) for the stellar photosphere.  
Although the photospheric output is easily calculated, 
the noise level for each star is more difficult to estimate.
At \70um, galactic cirrus contamination and extragalactic background
confusion are potentially limiting factors.
We screened the target stars for cirrus contamination with the IRSKY
tool at IPAC; 
interpolated fluxes from the low-resolution IRAS Sky Survey Atlas were
scaled to the smaller MIPS beam size based on the power spectrum of
the cirrus observed by IRAS \citep{gautier92}.
In addition to the noise contributed by the galactic cirrus, we also
set a minimum uncertainty for every image based on estimates of
extragalactic confusion \citep{dole03,Dole04}.

Beyond our primary criteria of spectral type F5-K5 and high expected
S/N, we apply several other secondary 
criteria.  Binaries whose point spread functions would significantly
overlap at \70um 
(separations less than 30\arcsec) are not considered.  
Also, to help populate different spectral type bins with similar
numbers of stars, a minimum photospheric \70um flux is set for each
spectral type bin: 20 mJy for F5-F9 stars, 10 mJy for G0-G4,
and 5 mJy for G5-K5.
There is no explicit selection based on stellar age or metallicity;
however, the \70um brightness and S/N thresholds are relaxed in
some cases to allow stars with well determined ages into the sample. 
There is no bias either for or against known planet-bearing stars.

\begin{figure}
\begin{center}
\includegraphics[width=4.7in,angle=-90]{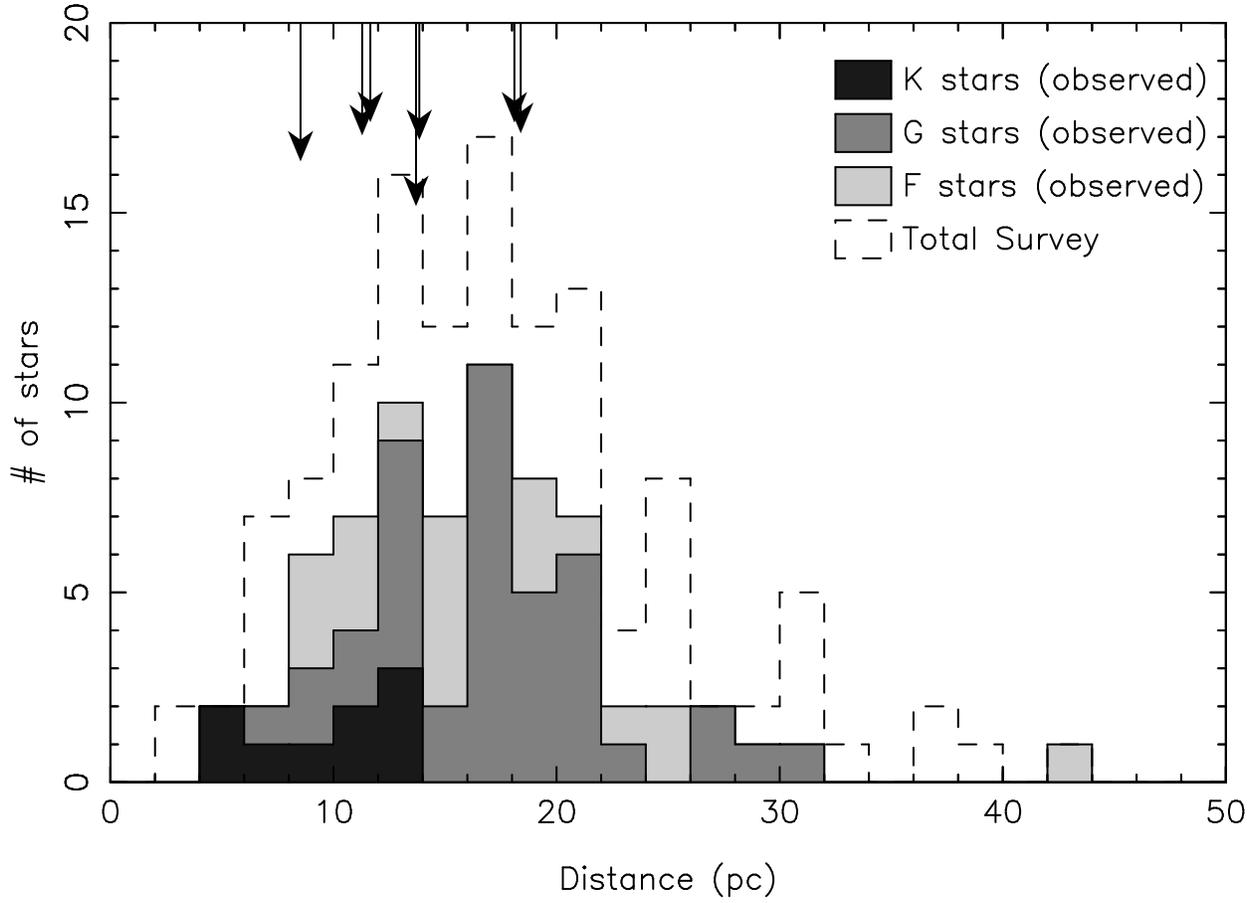} 
\end{center}
\caption{Distribution of stellar distances.
Each spectral type is shaded with a different color, as indicated in
the legend.
The distances of stars found to have \70um excess (see \S\ref{noisesec})
are flagged as arrows at the top of the plot. The length of the arrow
is an indicator of the strength of \70um excess.}
\label{dists}
\end{figure}

\begin{figure}
\begin{center}
\includegraphics[width=4.7in,angle=-90]{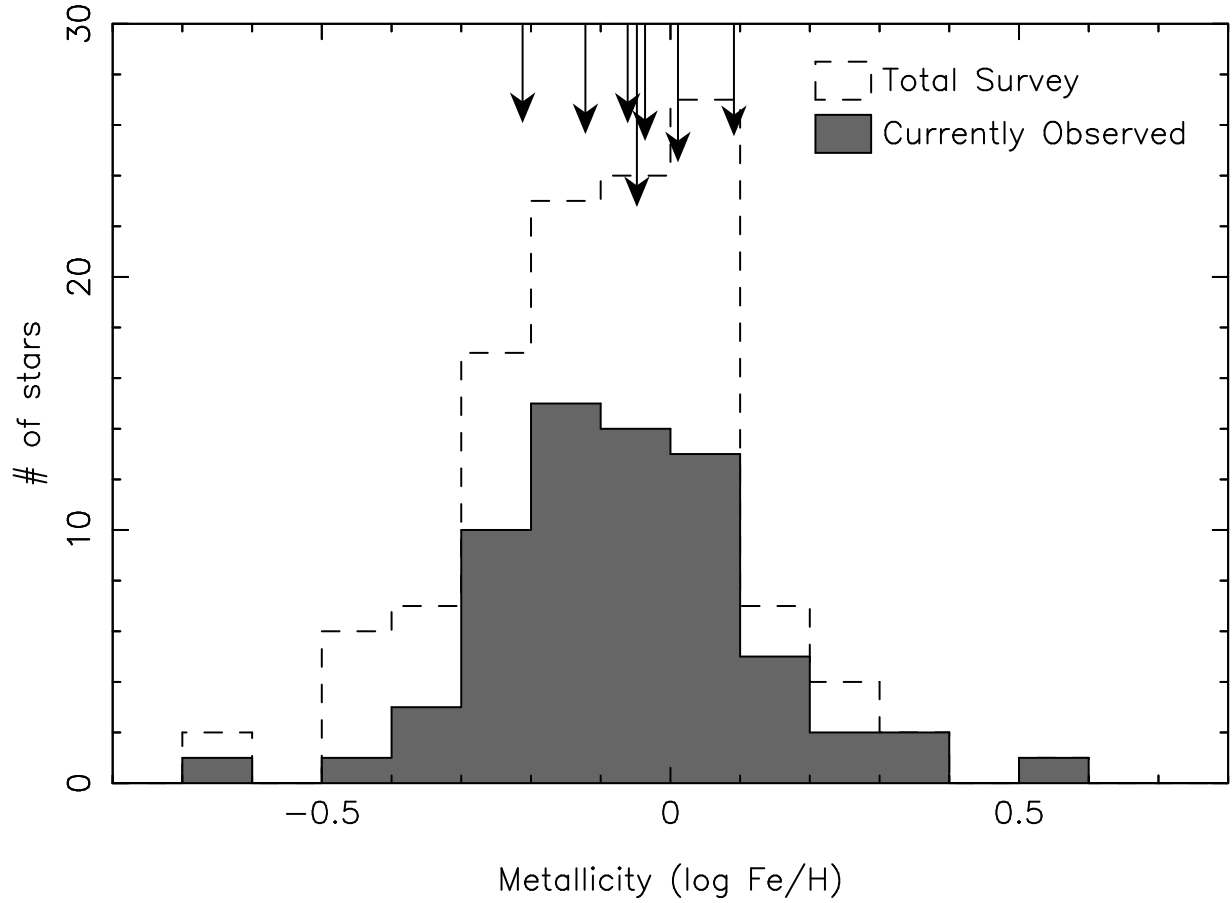} 
\end{center}
\caption{Distribution of stellar metallicities.
The metallicities of stars with \70um excess are flagged as arrows at
the top of the plot. The length of the arrow is an
indicator of the strength of \70um excess.  The detected excesses
are distributed uniformly over the stellar metallicities.}
\label{metals}
\end{figure}

The initial application of these criteria yields 131 stars.
Of these, four are observed by other guaranteed-time programs 
(see Table \ref{legacytable}).
This leaves 127 total stars, 69 of which have currently been observed
and are reported on here.
Binned by spectral type, the sample contains 
33 F5-F9 stars (20 observed), 46 G0-G4 stars (27 observed),
27 G5-G9 stars (13 observed), and 21 K0-K5 stars (9 observed).
Among these stars are 15 with known planets, 
of which 11 have been observed. 
Typical distances range from 10 to 20 pc, closer for K stars and
farther for earlier spectral types.
Fig.~\ref{dists} shows the distribution of stellar distances
as a function of spectral type, with filled 
histograms for the currently observed stars and a dotted, open
histogram for the eventual survey when complete.
Some basic parameters of the sample stars are
listed in Table~\ref{basictable}, most importantly age and
metallicity, which are also shown as histograms 
in Figs.~\ref{ages} and \ref{metals}. 


\section{Spitzer Observations}\label{observations}

\subsection{Data Reduction}

Our data reduction is based on the DAT software developed by the MIPS
instrument team \citep{Gordon05a}. 
For consistency, we use the same analysis tools and calibration
numbers as were adopted by \citet{Beichman05planets}.

At \24um, we carried out aperture photometry on reduced images as described in
\citet{Beichman05planets}. 
At \70um we used images processed beyond the standard DAT software 
to correct for time-dependent transients, corrections
which can significantly improve the sensitivity of the measurements
\citep{Gordon05b}.
Because the accuracy of the \70um data is limited by background noise,
rather than instrumental effects, a very small photometric aperture 
was used to maximize signal-to-noise -- just 1.5 pixels in radius.
With a 4 to 8 pixel radius sky annulus, this aperture size requires a
relatively large aperture correction of 1.79. 
The flux level is calibrated at 15,800 $\mu$Jy/arcsec$^2$/MIPS\_70\_unit, 
with a default color correction of 1.00
(MIPS\_70\_units are based on the ratio of the measured signal to the
stimulator flash signal).
Images were mosaiced from individual frames with half-pixel
subsampling.
For both the \24um and \70um data, neighboring point sources were
subtracted from the images before measuring the sky brightness.
With a telescope pointing accuracy of $<$1\arcsec\ \citep{werner04},
the stars are well centered within the chosen apertures; 
no centroiding is required.

\begin{figure}
\begin{center}
\includegraphics[width=4.7in,angle=-90]{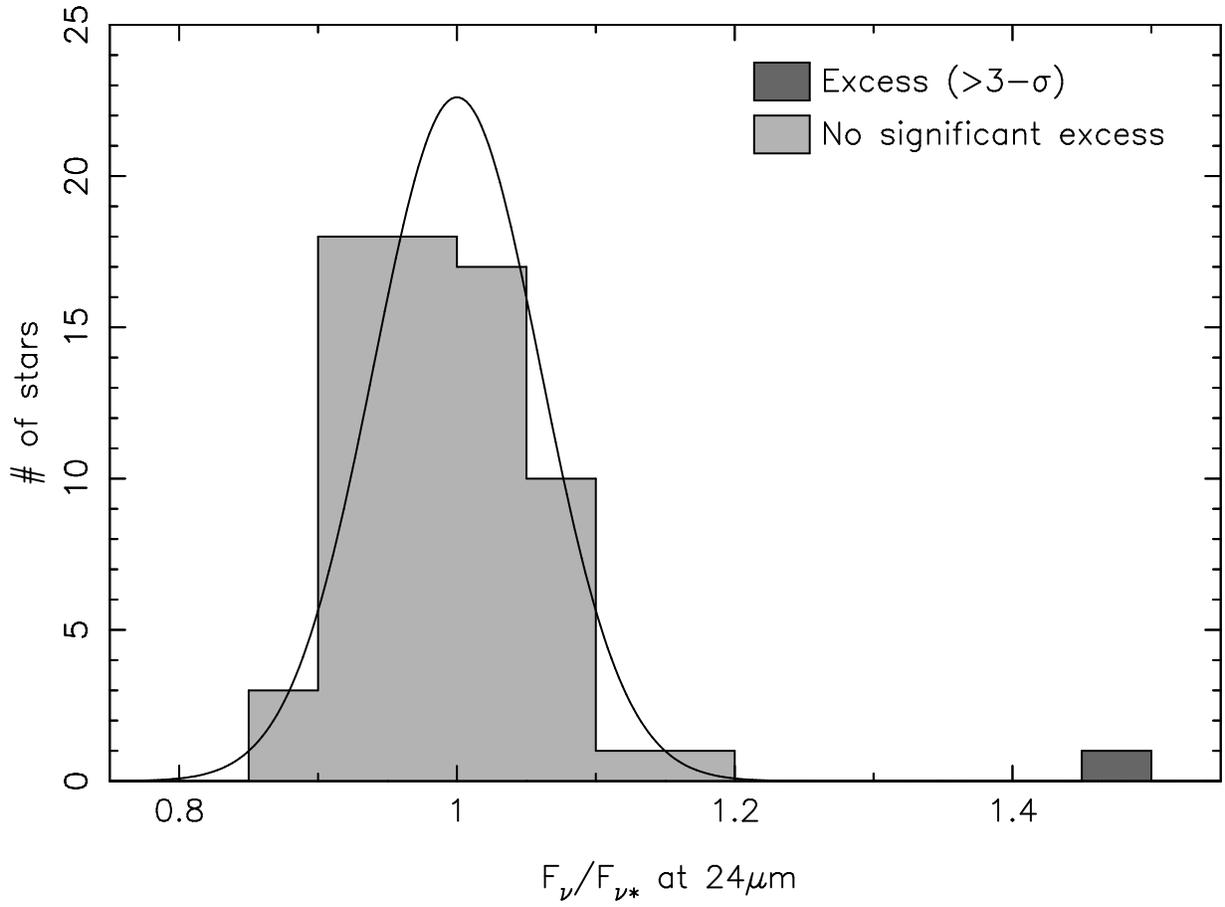} 
\end{center}
\caption{Distribution of \24um fluxes relative to the expected
photospheric values.
A Gaussian distribution with dispersion 0.06 is shown for comparison. 
One star (HD 69830) clearly stands out from the main population of
stars which do not have significant excess emission above their
stellar photospheres.} 
\label{f24k}
\end{figure}

To determine whether any of our target stars have an IR excess,
we compare the measured photometry to predicted photospheric levels
(\S\ref{kuruczsec}).
After excluding one outlying point (HD 69830), the 69 flux
measurements at \24um have an average $F_{\rm 
MIPS}/F_{\rm photosphere}$ of $0.99\,\pm\,0.01$; the agreement of the measured
fluxes with prediction is not surprising given that the present \spit\
calibration is based on similar stellar models. 
More importantly for determining the presence of any excess, the
dispersion of $F_{\rm MIPS}/F_{\rm photosphere}$ is 0.06 for this
sample (see Fig.~\ref{f24k}). 

At \70um, 55 out of 69 stars are detected with signal-to-noise ratio
$>3$.  This is in contrast with 
previous IR surveys of A-K stars with ISO, in which only half of the
stars without excess were detected \citep{Habing01}. While the
sensitivity of these \spit\ observations is roughly a factor of 10
better than previous data, \spit's accuracy is limited by
extragalactic source confusion and cirrus (see \S\ref{noisesec} below),
which will make it difficult to look for weak excesses around stars much
fainter than those discussed here. 

\begin{figure}
\begin{center}
\includegraphics[width=4.7in,angle=-90]{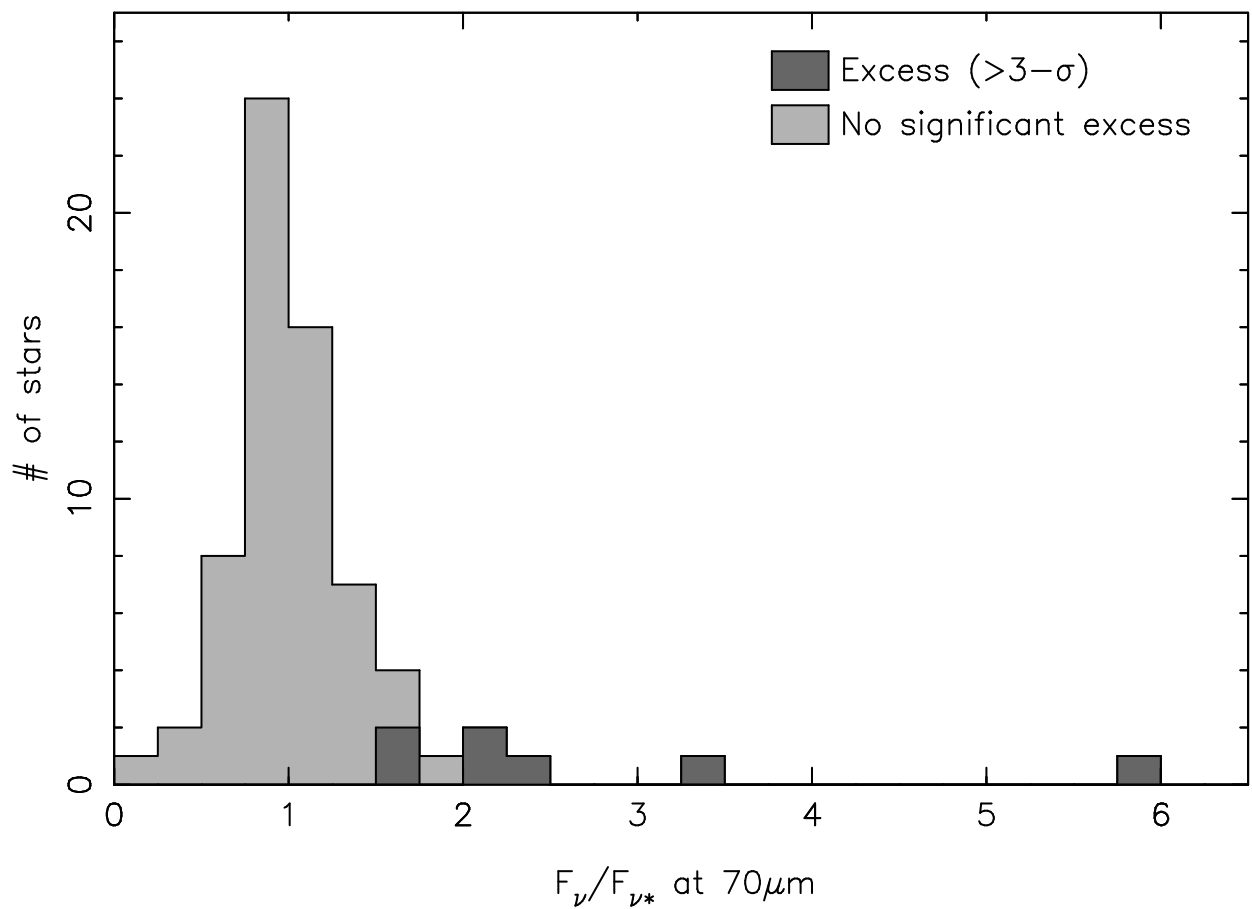} 
\end{center}
\caption{Distribution of \70um fluxes relative to the expected
photospheric values.
While most stars cluster around unity, where their flux is
photospheric, several stars show a high degree of excess emission.
Note that the distribution of excess stars is not completely
continuous;
whether a star has significant excess depends not only on
the measured flux ratio $F_{\nu}/F_{\nu \star}$, 
but also on the level of background noise
associated with each image.}
\label{f70k}
\end{figure}

The distribution of \70um flux densities relative to the expected
photospheric values is shown in Fig.~\ref{f70k}.
Unlike the tight distribution of flux ratios at \24um,
several stars have \70um flux density much higher than expected from the
stellar photosphere alone.   
Seven stars, with \70um fluxes densities from 1.7 to 5.8 times the expected
emission, are identified as having statistically
significant IR excess (see below).
Ignoring these stars with excesses and those with S/N $<3$,
the average ratio of MIPS flux to predicted photosphere is 
$F_{\rm MIPS}/F_{\rm photosphere} =1.08\,\pm\,0.03$, 
consistent with the present calibration precision.
The dispersion of $F_{\rm MIPS}/F_{\rm photosphere}$ relative to its mean
is 25\% in the \70um data (excluding the stars with excesses), considerably
higher than that in the \24um data (6\%).
The following section discusses in more
detail the noise levels within the \70um data.

\begin{figure}
\begin{center}
\includegraphics[width=4.8in,angle=-90]{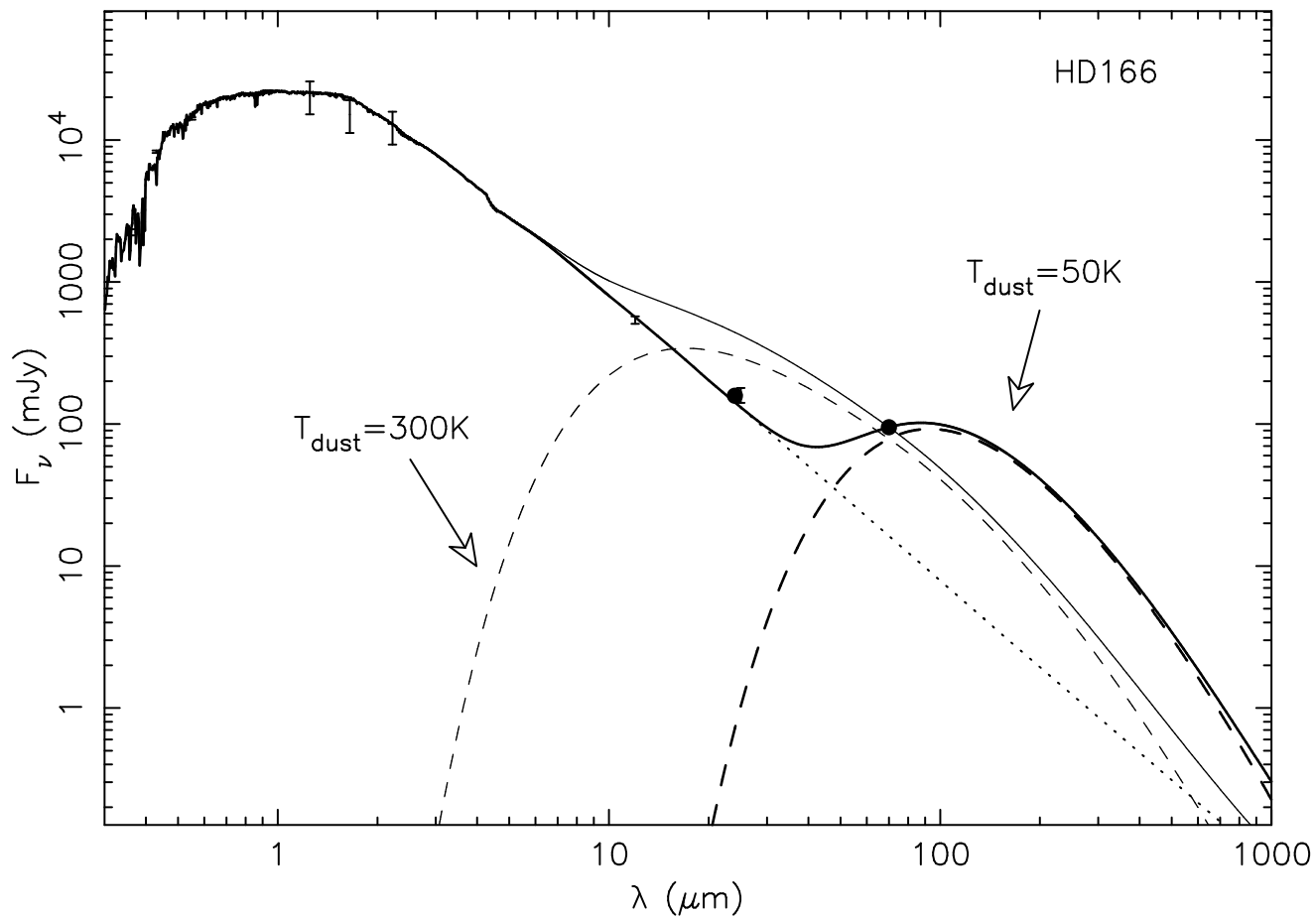} 
\end{center}
\caption{Spectral energy distribution for HD 166.
In addition to our 24 and \70um \spit\ data ({\it dark circles}), 
we also show optical, near-IR, and IRAS fluxes from the literature
({\it error bars}),
which are well fit by a stellar Kurucz model ({\it dotted line}).
Two possible dust temperatures are considered in order to fit the
\70um excess emission - 50 and 300~K ({\it dashed lines}). 
Only the cold, 50~K dust is consistent with the observed \24um flux;
hotter, 300~K dust is ruled out as the dominant source of IR excess.} 
\label{sedboth}
\end{figure}

Fig.~\ref{sedboth} shows an illustrative spectral energy distribution
for HD 166.  Published photometric fluxes for this star from visible
to infrared are well fit by a Kurucz stellar atmosphere 
\citep[{\it dotted line};][]{Castelli03,Kurucz03}.
The \spit/MIPS \24um flux is also well fit by the model atmosphere, but
the \70um emission is well above that expected from the stellar photosphere 
alone, requiring an additional component of emission due to dust.
Because there is only a single \70um measurement of IR excess, however, 
the SED can be fit by a range of dust temperatures and luminosities.  
A methodology for constraining the dust properties in these systems is
described in
\S\ref{ldsec}.

\subsection{Analysis of Background Noise}\label{noisesec}

An analysis of the noise levels in each field is required 
to assess whether the IR excesses are statistically significant. 
Many contributions to the overall error budget must be considered,
including those arising from stellar photosphere modeling
(\S\ref{kuruczsec}), instrument calibration, sky background variation, 
and photon detector noise.
For the \24um measurements, photon noise is negligible.
Even with the minimum integration time  (1 cycle of 3 
sec exposures = 42 sec), the sensitivity of MIPS is overwhelming;
our dimmest source could theoretically be detected in just a few
milliseconds.  
Also, the background noise is low: 
galactic cirrus is weak at this wavelength,
the zodiacal emission is relatively smooth across the field of view,
and the confusion limit for distant extragalactic sources is just
0.056 mJy \citep{Dole04}. 

Instead, for the \24um measurements, systematic errors dominate.  
The instrumental contribution to these errors is thought to be
very low: \24um observations of bright calibrator stars are stable with
1\% rms deviations over several months of observations \citep{Rieke04}. 
However, for photometry of fainter stars, dark latent images of bright
stars can result in larger errors because the star may be placed in a
dip in the flat-fielded photometric image frame.  
In addition to any uncertainty in the instrument calibration, the  
dispersion in $F_{\rm MIPS}/F_{\rm photosphere}$ includes errors
in the photosphere extrapolation as well as the effects of source
variability.  
The fitting of the photosphere can be as precise as 2\% when the best
2MASS $K_s$ band photometry is available \citep{skrutskie00}, but for stars
brighter than $K_s \simeq 4$ mag, 2MASS data are less accurate and
lower precision near-IR data and/or shorter wavelength observations
must be relied upon.  
Extrapolation from visible data places considerably more weight
on the photospheric models, increasing the uncertainty in
the predicted photospheric levels at \24um and \70um. 
We expect that stellar photosphere fitting errors
and flat field uncertainties due to latent images 
are the greatest contributors to the overall error budget.
The net photometric accuracy is currently $\sim$6\%,
as seen in the dispersion in Fig.~\ref{f24k}.
Detections of excess at \24um (at the 3-$\sigma$ level) require
measured fluxes at least 20\% above the stellar photosphere
(about 1000 times the Solar System's \24um excess flux ratio).

While systematic errors dominate at \24um, for \70um data
pixel-to-pixel sky variability becomes a major contributor to the
overall uncertainty.
This sky variation is a combination of detector/photon noise along
with real fluctuations in the background flux.  
This background, a combination of galactic cirrus and extragalactic
confusion, creates a noise floor that cannot be
improved with increased integration time.  
To minimize this problem, the FGK target stars are chosen from
areas of low galactic cirrus, as estimated from the IRAS Sky Survey Atlas
\citep{ipac94}. 
The confusion limit for extragalactic background sources, however, is
unavoidable and sets a strict lower limit for the sky noise at \70um.

We determine the pixel-to-pixel noise in each field by convolving the
background image with the same top-hat aperture used for photometry
(1.5 pixel radius) and then by 
calculating the dispersion within these background measurements.
The region within 3 pixels of the target is excluded.
The error on the mean noise is proportional to the square root of the
number of contributing apertures.
Based on this overall measured noise, we find the S/N for 
each star, as listed in Table~\ref{resultstable}.
The median observed S/N for our target stars is $\sim$6, excluding the
sources identified as having excess emission. 

The measured noise (also listed in Table~\ref{resultstable}) can be
compared to that expected from extragalactic confusion. 
\citet{Dole04a,Dole04} find a 5-$\sigma$ confusion limit of 3.2 mJy 
by extrapolating \spit\ source counts of bright objects down
to fainter fluxes. 
In our sample, the lowest (1-$\sigma$) 
noise levels observed toward stars located in clean portions of the
sky are $\sim$2 mJy, somewhat worse than Dole et al.'s best-case limit.
This difference is attributable to the larger effective beam size 
used for our photometry/noise calculations, 
and to confusion noise in the limited sky area in our images.
On top of the confusion limit, a few sources have higher noise values
due to galactic cirrus and/or 
detector performance somewhat worse than typical. 

\begin{figure}
\begin{center}
\includegraphics[width=4.8in,angle=-90]{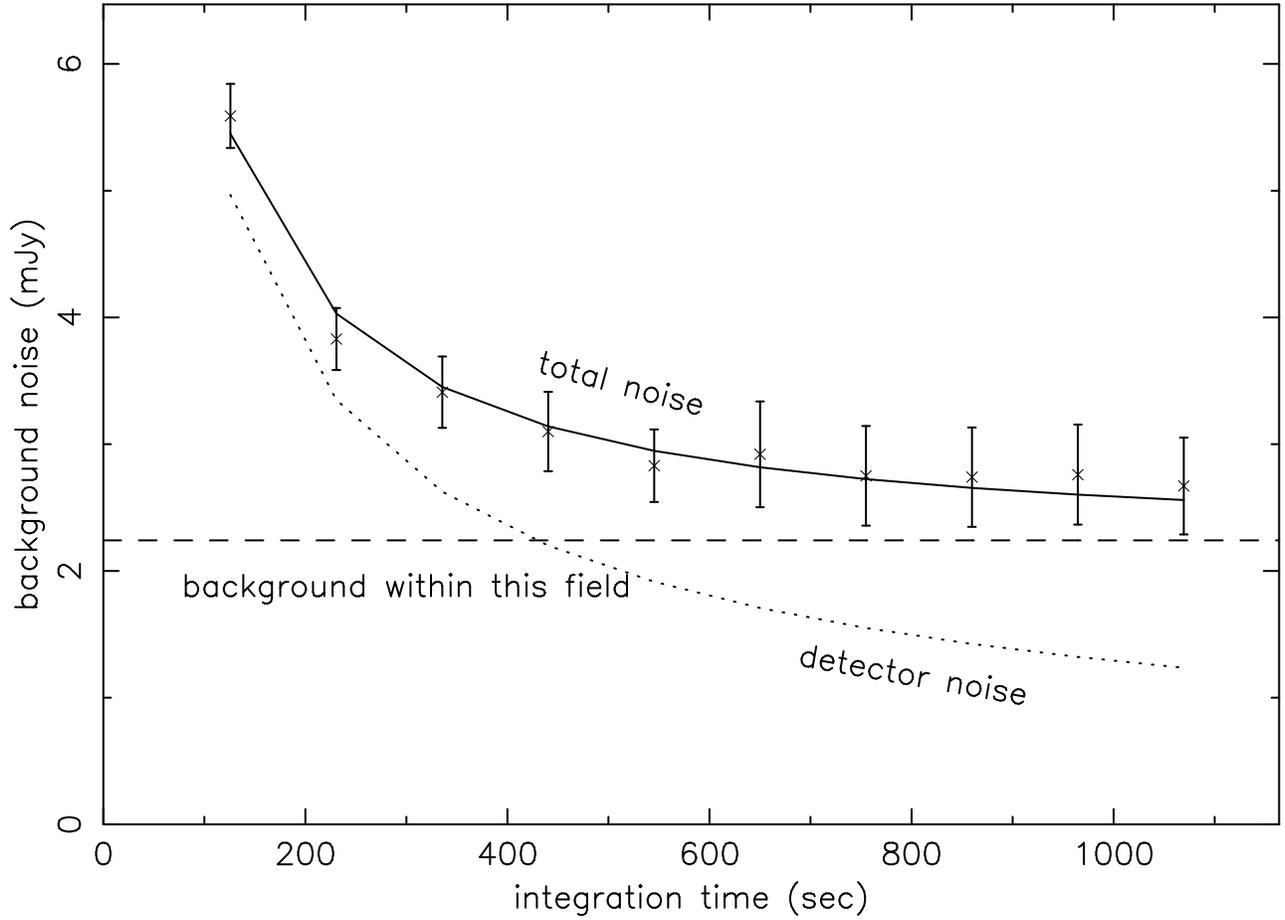} 
\end{center}
\caption{\70um background noise as a function of MIPS integration time
for the field surrounding HD 62613.  Measurements within the MIPS
frames from 1 to 10 cycles ({\it crosses}) are fit by a combination of
detector noise ({\it dotted}) and underlying background ({\it
dashed}).  While photon detector noise decreases with integration
time, the background, 
a combination of galactic cirrus and extragalactic confusion, does
not. } 
\label{noisefits}
\end{figure}

The sky fluctuations in each field are a combination of detector noise plus
real background variations.
When the total observation can be separated into individual snapshots
with shorter integration time (i.e.\ when there are multiple observing
cycles), we can isolate the two sources of noise.
We create several images at each integration time by separating the
individual cycles and then adding chains of them together of various lengths.
In each case, the noise is assumed to come from two terms added in
quadrature, one constant and one declining with time.
Specifically, the noise is fit to a function 
$\sqrt{B^2 + (D t_{\rm int}^x)^2}$ 
where $B$ is the constant background, $D$ is the strength of the
detector noise for 1 observing cycle, and $t_{\rm int}$ is the
integration time (in cycles). 
Fig.~\ref{noisefits} shows the resulting fit for HD 62613, 
a star observed for 10 cycles. 

Naively, one would assume that detector noise drops off as the square
root of integration time (i.e.\ $x=-1/2$).
In practice, however, we find a typical time dependence of $x \simeq
-0.6$. In other words, the noise drops off {\it faster} than expected.
This surprising result follows from our method of data processing,
which improves as more images are included in the analysis.
The time-filtering routines have an optimal filtering window of 3-4
observing cycles \citep{Gordon05b}, such that four cycles of integration time
($\simeq 400$ sec) result in less than half the detector noise of a
single cycle ($\simeq 100$ sec).

\begin{figure}
\begin{center}
\includegraphics[width=4.8in,angle=-90]{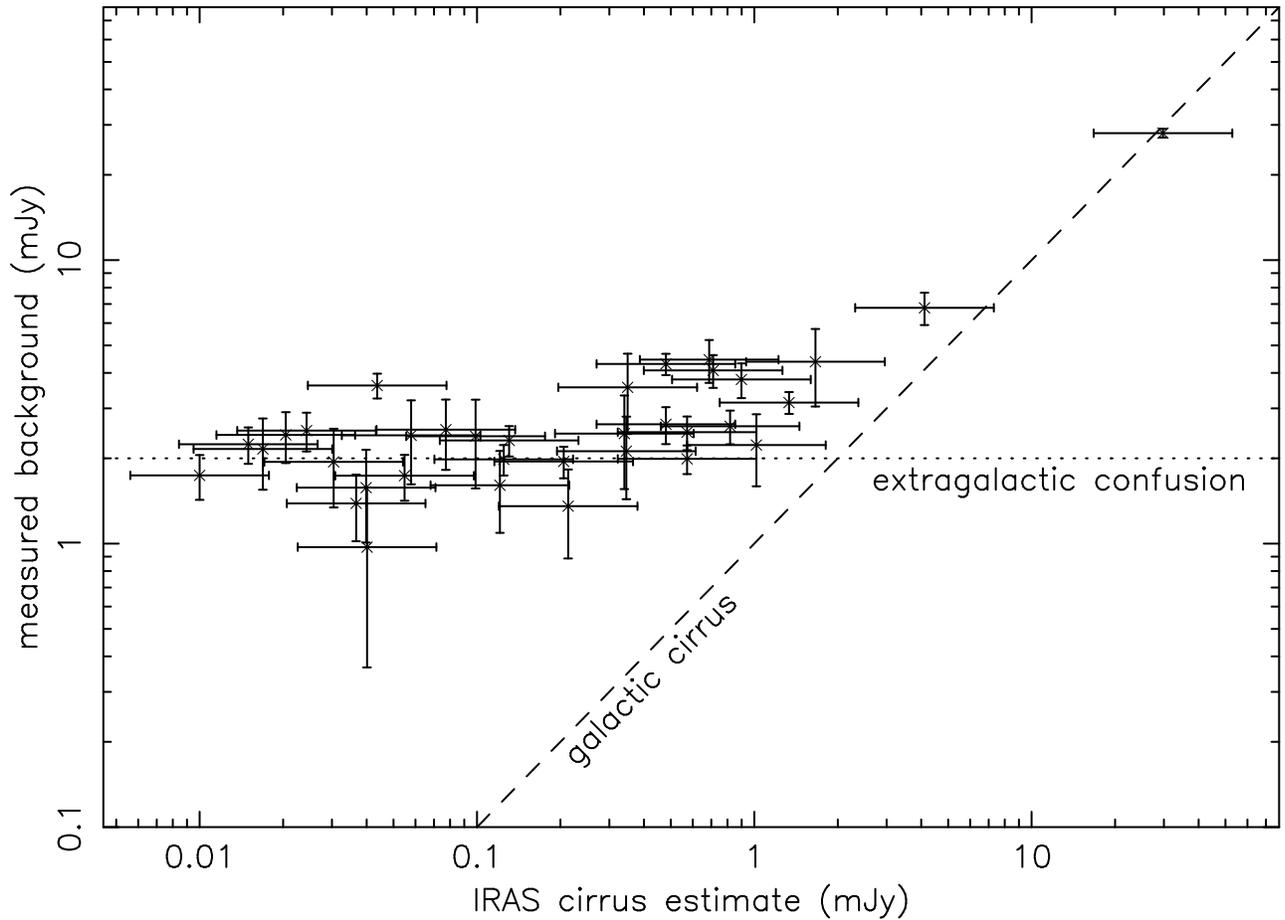} 
\end{center}
\caption{Measured background noise level within \70um MIPS images
compared with the estimated cirrus background based on IRAS data.
The MIPS noise here is from background only; detector noise has been
subtracted to the extent given by the error bars.
Only shown are those stars with enough images (at least 4 observing
cycles) to disentangle the detector noise from the real sky background.
The overall trend of the noise is well fit by the
IRAS estimate for galactic cirrus ({\it dashed line}) combined with a
$\sim$2 mJy lower limit from extragalactic confusion 
({\it dotted}).}
\label{irasnoise}
\end{figure}

Fig.~\ref{irasnoise} shows how the background in our MIPS data
compares to IRAS-based predictions.  
For the single star with a very high level of cirrus contamination
(HD168442, the rightmost point in Fig.~\ref{irasnoise}), 
the IRAS noise level agrees well with that in the higher resolution
MIPS field.   Because the stars in this sample
were pre-selected from regions of low cirrus contamination, however, 
the majority are not dominated by cirrus, and instead have background
noise levels close to the extragalactic confusion limit.
Note that the confusion limit here ($\sim$2 mJy) depends on our method
of photometry (aperture size, shape, sky subtraction) and does not 
necessarily reflect the intrinsic properties of the instrument and the
observed fields. 

Finally, we consider any systematic errors.
While we can directly examine the overall background noise in each of
our \70um images, the systematic uncertainties are more difficult to evaluate.
Repeated measurements of bright standards have rms scatter of $\sim$5\%. 
The photospheric extrapolations may contribute 6\% (judging from \24um). 
The detectors may also have a low level of uncorrected nonlinearity. 
We assume that the systematic errors in the \70um data are 15\% 
of the stellar flux, about twice the dispersion in the \24um data. 

Adding all of the noise sources (photon noise, sky background, model
fitting, and residual calibration issues) together in quadrature gives
us a final noise estimate for each \70um target.
In Table \ref{resultstable} we list these noise levels, along with the
measured and the photospheric fluxes, for each observed star.  
We use these noise estimates to calculate $\chi_{70}$, the statistical
significance of any IR excess
\begin{equation}\label{chi70eq}
\chi_{70} \equiv \frac{F_{70} - F_{\star}}{N_{70}}
\end{equation}
where $F_{70}$ is the measured flux, $F_{\star}$ is the expected
stellar flux, and $N_{70}$ is the noise level, all at \70um.
Based on this criterion, 
we find that 7 out of 69 stars have a 3-$\sigma$ or greater excess at
\70um: HD 166, HD 33262, HD 72905, HD 76151, HD 115617, HD 117176, 
and HD 206860. 
Of the remaining stars, 3-$\sigma$ upper limits on any excess flux
vary from star to star, but are generally comparable to the stellar
flux at \70um (the median upper limit is 0.8 $F_{\star}$).

While a strict 3-$\sigma$ cutoff is useful for identifying the stars
most likely to have IR excess at \70um, several other stars below this
limit may also harbor similar amounts of dust.
HD~117043, for example, has a \70um flux density twice that expected
from the stellar photosphere.  A relatively dim source, this potential
IR excess is not significant at the 3-$\sigma$ level ($\chi_{70}= 2.4$), 
but is corroborated by a similarly high \24um flux (11\% above photospheric).
Similarly, \spit/IRS spectra can provide additional evidence for
borderline cases.
In all three cases where spectra have been obtained for stars with
$>$3-$\sigma$ \70um excesses (HD 72905, HD 76151, and HD 206860) 
each spectrum contains clear evidence of a small excess 
at its longest wavelengths \citep[from $\sim$25 to 35 \um;][]{Beichman05IRS}.
Another star, HD 7570, with only 1.8-$\sigma$ significant excess at
\70um, has a similar upturn in its spectra, suggesting that its
moderately high level of \70um flux ($F_{\rm MIPS}/F_{\star} = 1.5$) is in
fact excess emission.

\section{Properties of the Detected Dust}\label{ldsec}

Our detections of IR excess provide only 
limited information about the properties of the dust in each system.
In principle, each observed wavelength translates to a characteristic
radial-dependent temperature and thus can tell us about a particular
region of the dust disk.
The exact location of dust at a given temperature depends on the
stellar luminosity and on the grain emissivities.
In general, though, dust in the inner system ($\lapp$10 AU) has
temperatures $\gapp$150 K and radiates strongly at \24um.
The emission of dust at Kuiper Belt-like distances, with temperatures
$\sim$50 K, peaks closer to \70um.
For our \70um dust detections, the lack of \24um excess limits the
amount of material in the inner system.
In these cases, \24um measurements provide an upper
limit on the dust temperature (as in 
Fig.~\ref{sedboth}), while sub-mm observations would set a lower limit. 
Because we usually have no information longward of \70um, however,
only an upper limit on the temperature (or an inner 
limit on the dust's orbital location) can be derived.

If a single dust temperature is assumed, the observed flux can
be translated into the total dust disk luminosity relative 
to its parent star. 
For disks with detections of IR excess, a minimum dust
luminosity can be calculated.  On the Rayleigh-Jeans tail of 
the stellar blackbody curve, the ratio of dust to stellar fluxes is
\begin{equation}\label{bbeq}
\frac{F_{{\rm dust}}}{F_{\star}} = 
\frac{L_{\rm dust}}{L_{\star}} \;
\frac{h\nu T_{\star}^3}{kT_{\rm dust}^4 (e^{h\nu/kT_{\rm dust}} -1)}
\end{equation}
The minimum disk luminosity as a function of \70um dust flux can be
obtained by setting the emission peak at \70um
(or, equivalently, $T_{\rm dust} = 52.5$K) : 
\begin{equation}\label{ldeq}
\frac{L_{\rm dust}}{L_{\star}}({\rm minimum}) = 10^{-5} 
\; \left(\frac{5600 \; {\rm K}}{T_{\star}}\right)^3
\; \frac{F_{70, {\rm dust}}}{F_{70, \star}}
\end{equation}

\begin{figure}
\begin{center}
\includegraphics[width=5.2in]{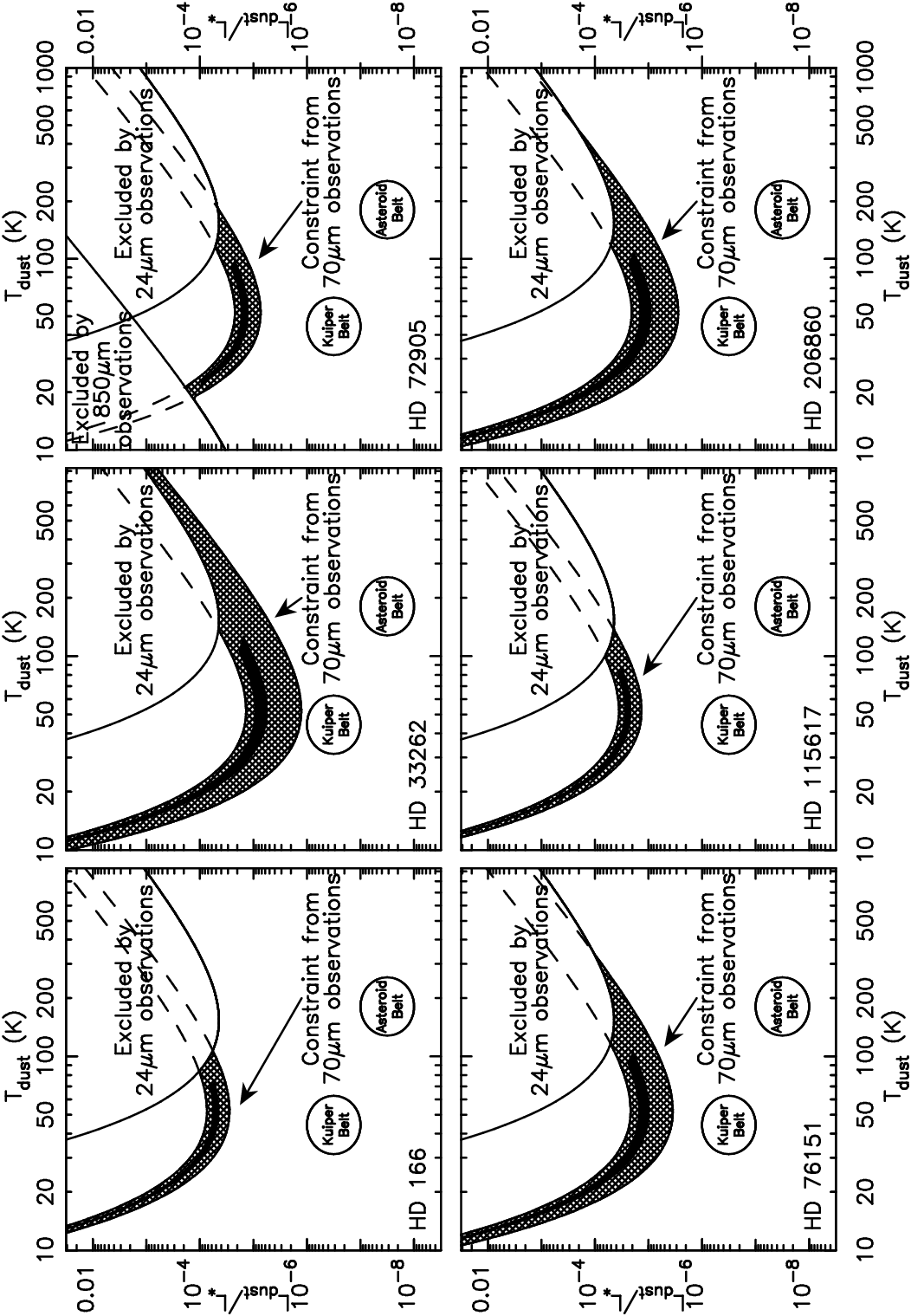} 
\end{center}
\caption{Constraints on the temperature and total luminosity of 
the dust around six stars with \70um excess emission.
The upper right of each panel (bright, hot emission) 
is ruled out by the \24um upper limits.
For HD 72905 (upper right panel) the upper left of the plot (bright, cold
emission) is also ruled out by sub-mm observations \citep{Greaves05}.
Based on the measured \70um excess, possible dust
temperatures and luminosities are shown as shaded regions 
(cross-hatched for 3-$\sigma$ error limits; solid-filled  
for 1-$\sigma$ limits). 
The approximate characteristics of the asteroid and Kuiper
belts are shown for comparison.}
\label{ldlstar2}
\end{figure}

Based on this equation, a minimum $\ld$ is calculated for 
each of our target stars identified as having IR excess 
(see Table \ref{resultstable}). 
The disk luminosity, however, could be greater than this value,
depending on the dust temperature.
In particular, a much larger amount of radiation
could be emitted at unobserved sub-mm wavelengths.
Fig.~\ref{ldlstar2} shows the overall constraints on $\ld$ as a
function of $T_{\rm dust}$ for six stars identified as having excess
\70um emission.   The lines in this figure are 3-$\sigma$ limits to
the observed 24 and \70um fluxes, while the filled, dark region
corresponds to 1-$\sigma$ limits.
The lack of excess emission at \24um excludes the upper right region
of each plot (i.e.\ bright, hot emission) and typically constrains the
dust temperature to be $\lapp$100 K at the 1-$\sigma$ level 
({\it solid-filled region}).   

Sub-mm observations are critical for constraining 
the dust properties beyond an upper bound for temperature and a lower
limit for luminosity.
For most of our stars with IR excess, large amounts of cold dust 
emitting at longer wavelengths cannot be ruled out.  
HD 72905, however, has been observed at 850 \um,
with a measured flux of 1.1$\,\pm\,$1.2 mJy \citep{Greaves05}.
Cold, very bright emission is excluded.
Note that for the sub-mm flux, it is no longer appropriate to assume
blackbody emission;
as the wavelength of the emitted radiation becomes long compared to
$2 \pi a_{grain}$, the effective grain absorption cross-section begins
to fall off as $\lambda^{-1}$ to $\lambda^{-2}$ \citep{Draine84,Wyatt02}.
In order to calculate the most conservative limit on $\ld$, we assume
that the grains are small enough such that their emissivity drops off
as $\lambda^{-2}$ for radiation longward of 100 \um.
With the inclusion of this sub-mm limit, the dust temperature and
luminosity for HD72905 are bounded by 1-$\sigma$ limits ({\it
solid-filled region in Fig.~\ref{ldlstar2}}) of $T_{\rm dust} \approx
20$-100 K and $\ld \approx 10^{-4.9}$-$10^{-4.1}$, i.e., within about
a factor of six of the lower limit from 70$\mu$m data alone.  

$T_{\rm dust}$ is meant to signify the typical emitting temperature
for the dust; in reality some range of temperatures will be found in
any given system.
The approximate characteristics of the dust in the solar system, for
example, have been included in Fig.~\ref{ldlstar2}, where the Kuiper
and asteroid belts are shown as separate regions with discrete
temperatures. 
There is growing evidence for multiple-component dust disks around
other stars as well. 
Resolved images of the bright disk around Fomalhaut
\citep{stapelfeldt04} 
show that 24 and \70um
emission can have markedly different spatial distributions.  
Observations of $\epsilon$ Eri's disk \citep{megeath05} similarly find
70 and 850 \um\ emission coming from completely distinct regions.
For the unresolved sources considered here,
the strong excess emission that we observe at \70um is clearly
due to dust with colder temperatures than the asteroid belt,
but a lower level of warm dust cannot be ruled out.
In fact, follow-up IRS spectra suggest that small levels of warm dust
orbiting inside of the dominant outer dust may be common in these
systems \citep{Beichman05IRS}.

Even with full spectral and spatial coverage, it is difficult to determine
the amount of dust responsible for the excess emission
around these stars, let alone the overall mass of larger bodies that
create the dust.
Given the dust luminosity and temperature, the total 
cross-sectional area of the dust, $A_d$, is 
\begin{equation}
A_d =
\frac{\pi R_{\star}^2}{\epsir}
\, \frac{L_{\rm dust}}{L_{\star}}
\, \frac{T_{\star}^4}{T_{\rm dust}^4}
\label{area}
\end{equation}
where $\epsir$ is the median grain emissivity over all wavelengths. 
For debris disks, this emissivity is observed to drop off as
$\sim\lambda^{-1}$ at sub-mm wavelengths \citep{dent00},
consistent with the particle size distribution expected from a
collisional cascade.
In general, though, the magnitude of the grain emissivity is uncertain.
Here, we assume that $\epsir$ can be as high as unity if the dust is
warm, but might be several orders of magnitude lower for cold dust.
Even without this uncertainty in $\epsir$,
the dust area given by Eq.~\ref{area} is not well determined.
Among our stars with IR excess, HD 72905
has the best constraints on disk brightness and temperature, 
yet the dust cross section can still range anywhere
from $10^{24}$ to $10^{29}$ cm$^2$ (1-$\sigma$ limits).
If the dust consists solely of micron-sized grains, this area
corresponds to a dust mass of $\sim$10$^{-7}$-$10^{-2} \MEarth$.
The total mass of the debris disk, which tends to be dominated by the most
massive objects, is even less well constrained.   
Under the assumption that the number of particles of a given size,
$\rg$, follows the equilibrium size distribution $dn/d\rg \propto
\rg^{-3.5}$ \citep[e.g.][]{dohnanyi68},
the total mass can be estimated as a
function of the largest planetesimal size, $\rgmax$.
For HD 72905, the disk mass is 
$\sim$10$^{-2}$-10$^3$ $\MEarth \sqrt{\rgmax/10 \, {\rm km}}$.
The range of values in these mass estimates reflects the
uncertainty in the dust location within these unresolved images
(anywhere from $\sim$10 to 100's of AU for the dominant component).
The largest disks consistent with the SED observations can be 
ruled out by the lack of emission extended beyond the telescope PSF 
At HD 72905's distance of 14.3 pc, a 300 AU diameter disk,
subtending an angle of $\sim$2 instrument pixels, would be
clearly extended in the \70um image.
(The telescope's FWHM at \70um is 18\arcsec.)

Although the dust temperature is generally not well determined by a
single measurement of excess at \70um, the emission of HD 72905 in
particular has been further constrained on the short end by IRS
observations \citep{Beichman05IRS}. 
The upturn in the spectrum longward of $\sim$25 \um\ provides a very
sensitive measure of the maximum dust temperature.
From the combined spectral and photometric data,
\citet{Beichman05IRS} estimate dust temperatures and masses for a
variety of grain properties.  For small 0.25 \um\ grains,
they find dust temperatures ranging from 35 to 55 K, corresponding to 
a dust mass of $10^{-4} \MEarth$,
consistent with the above estimates.

For stars with no detected emission, 3-$\sigma$ upper bounds on the
\70um fluxes lead to upper limits on $\ld$ as low as 
as few times $10^{-6}$, assuming a dust temperature of $\sim$50 K
(see Table \ref{resultstable}).
Although we cannot rule out cold dust at $\gapp$100 AU,
we are placing constraints on dust at Kuiper Belt
distances at $\sim$10-100 times the level of dust in our solar system. 
The constraint on asteroid belt-type dust is less stringent, at
$\sim$1000 times our zodiacal emission.
Many of our stars are bright ($>$30 mJy at \70um) and in regions of
low background (a few mJy), such that the greatest source of error
in their \70um flux is due to the overall calibration uncertainty of
$\sim$15\%. 
This uncertainty sets a threshold for minimum detectable $\ld$ at 
$5 \times 10^{-6} \, (\Tstar/5600 \, K)^{-3}$ (from Eq.~\ref{ldeq}).

\section{Correlation of Excess with System Parameters}\label{parameters}

To understand the origin of any excess,
we now consider the properties of the sample stars and how they
correlate with excess detection.
Specifically, we examine the correlation with three variables: 
1) age,
2) metallicity, and
3) spectral type.
These parameters are listed for each star in Table~\ref{basictable}. 

\subsection{Age}

Stellar youth is already well established as a primary indicator for
excess IR emission \citep{spangler01, Rieke05}.
This connection is often
interpreted as a continual decline in disk mass with time.
Young stars lose their protostellar disks relatively quickly,
transitioning from gaseous disks into less massive debris disks on
time scales of $\sim$3~Myr \citep{haisch01}.
While there is a correlation between stellar age and disk
emission, the assumption that all debris disks gradually grind
down into weak disks like the Sun's is contradicted by
observations of old stars with IR excess \citep{Habing01,decin00,Rieke05}.
Strong collision events may be able to increase the dust emission,
even at late times.  

Unfortunately, there is no generally reliable age indicator for stars
as old as those in our sample.   
Age estimates for our target stars generally span at least a factor of
two, highlighting the difficulty in determining the ages of mature, 
main sequence stars.    
Whenever possible we adopt ages from the compilation of
\citet{Wright04}, which provides a uniform 
tabulation for 1200 stars based on Ca II H\&K line strengths.
Otherwise an average of literature values is calculated.
In addition to listing this age estimate for each of our target stars,
Table~\ref{basictable} also gives the maximum and minimum age
found in the literature (for stars with more than one age estimate).

\begin{figure}
\begin{center}
\includegraphics[width=4.7in,angle=-90]{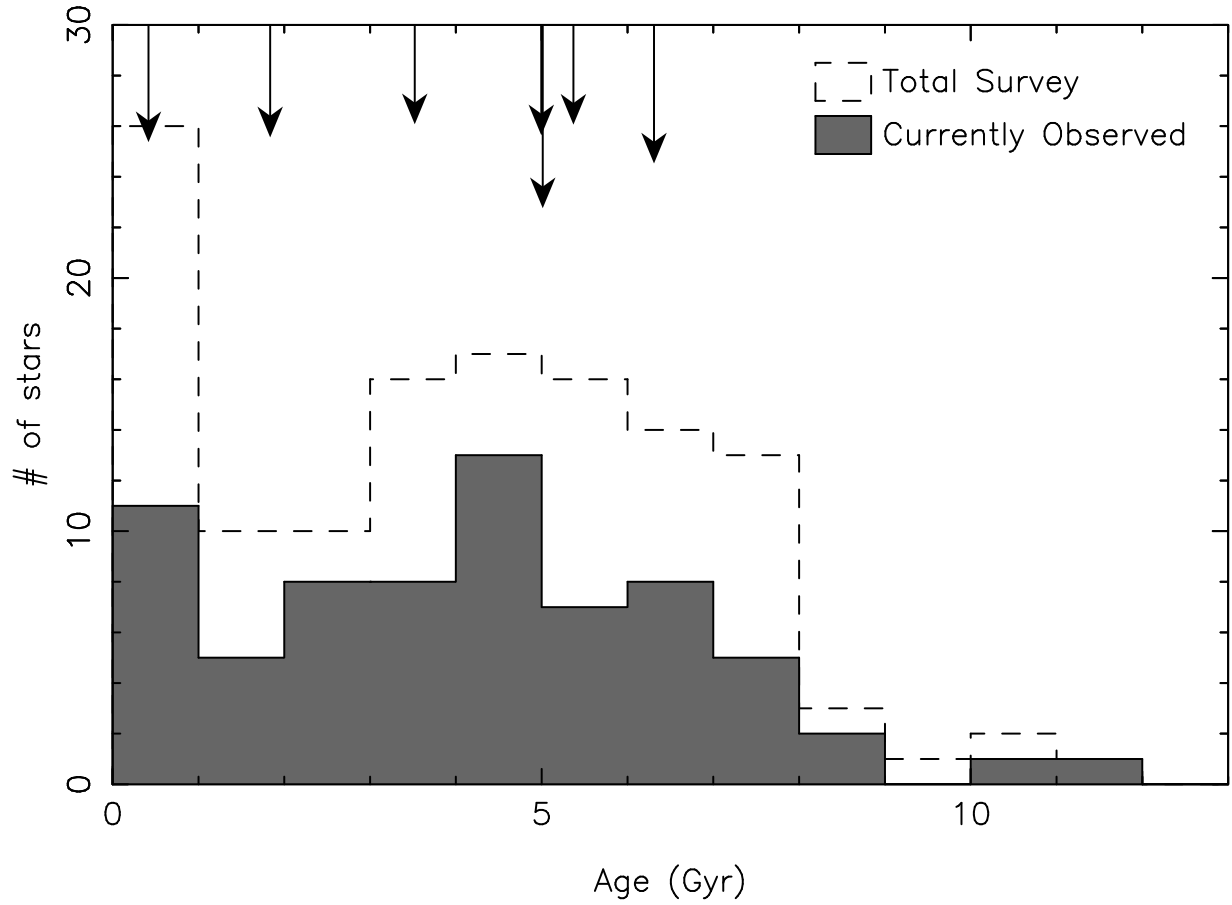} 
\end{center}
\caption{Distribution of stellar ages.
The ages of stars with \70um excess are flagged as arrows at the top
of the plot. The length of the arrow is an
indicator of the strength of \70um excess.  There is no strong
correlation between the detection of an excess and the stellar age.}
\label{ages}
\end{figure}

Fig.~\ref{ages} shows the resultant histogram of stellar ages.
Although our target selection criteria do not explicitly
discriminate based on stellar age, 
young stars are not well represented in our sample due to
their infrequent occurrence within $\sim$25 pc of the Sun.
Therefore, our data cannot probe the rapid ($\sim$100 Myr)
initial decline seen for samples of young stars, but instead are
sensitive to any trends that occur over Gyr time scales.

The ages of the stars with excess are marked with arrows in Fig.~\ref{ages}.  
Most of the stars with excess are much older than a billion years;  
the one exception is HD 72905 which has a young age estimate
(0.42~Gyr) based on its relatively high stellar activity.
The average age of the stars with \70um excess is nearly identical to
the average of the sample as a whole ($\sim$4 Gyr). 
We find no statistically significant correlation between age and excess.

\subsection{Metallicity}

The relationship of disk properties to the metallicity of 
the parent star is particularly important for understanding the formation
and evolution of debris disks and, more generally, of larger planets.
One might expect that the formation of objects composed of metals (i.e.\ dust,
planetesimals, and terrestrial planets) will be strongly correlated
with stellar metallicity.  Gas giant planets, if their formation is
preceded by the formation of a large solid core 
\citep[e.g.][]{pollack96}, should also depend on the amount of solid
material available in the protostellar disk.
Alternately, if gas giants form via direct gravitational collapse
of the disk \citep[e.g.][]{boss04}, planet formation would only depend
on metallicity through less important opacity effects.

In fact, there is a well known correlation between extrasolar
gas giant planets and host star metallicity
\citep{gonzales97,santos01}. 
In particular, \citet{fischer05} find that the probability of
harboring a radial-velocity detected planet increases as the square of 
the metallicity.
However, there is as yet no evidence for a similar correlation
between dust and metallicity.
\citet{Greaves05b} even find an anti-correlation
between metallicity and dusty debris at sub-mm wavelengths. 
As an example, the $\sim$10 Gyr-old star $\tau$ Ceti 
has strong excess emission in both sub-mm \citep{GreavesTauCeti}
and infrared wavelengths (Table~\ref{legacytable}), 
despite having only a third the metals of the Sun.

To look for any positive or negative correlation between
metallicity and  IR excess within our observed stars,
we have collected metallicity data from the literature for our FGK targets.
The majority of [Fe/H] values are derived from
spectroscopic analysis; a few are from narrow-band filter photometry. 
While for some stars as many as seven independent values for [Fe/H] are
available, no abundance information is available for five stars.
Table~\ref{basictable} lists the number of independent [Fe/H]
estimates, their average, and the r.m.s. scatter for each star. 

Fig.~\ref{metals} shows a histogram of these metallicity values, 
which range from -0.5 to +0.5 dex with a mean value just below solar. 
The stars with IR excess are again identified with vertical arrows.
In spite of any expectations,
there is no evidence for higher metallicity resulting in a greater
amount of IR emitting dust. 
The average [Fe/H] is -0.07$\,\pm\,$0.02 for the observed stars 
and -0.05$\,\pm\,$0.04 for the stars with excess -- a small and
insignificant difference.
The correlation coefficient, $r$, between [Fe/H] and IR excess is
0.02$\,\pm\,$0.12. 
The strong type of relationship found between gas giant planets and
metallicity would have resulted in a much stronger correlation
($r$=0.38$\,\pm\,$0.13) and can be confidently ruled out.

The lack of correlation between excess and metallicity is somewhat
surprising given the strong correlation between planets and
metallicity and the preliminary 
correlation that we have found between planets and excess
\citep{Beichman05planets}. 
Our sample here, however, contains relatively few planet-bearing stars
(11 out of 69).  While these stars do have higher metallicity,
only 1 out of 11 has an IR excess, resulting in a detection
rate very similar to the non-planet stars and causing no net increase
in the average metallicity of excess stars.
If all of the planet-bearing stars described in
\citet{Beichman05planets} are included within this sample,
the correlation coefficient increases somewhat, but still not to a
significant level.

While the lack of a metallicity-excess correlation may be surprising,
there are several possible explanations.
The formation of giant planets, which do have a strong metallicity
correlation, requires a very massive protoplanetary disk.  
The disk that developed into the solar system, for example, 
originally contained more than $100 \MEarth$ of solid material,
based on the composition of the planets today \citep{hayashi85}.
Our Kuiper Belt is much smaller, currently containing only 
a few percent of $\MEarth$ \citep{bernstein04}.
In fact, very little mass is needed to produce the dust responsible
for the observed IR excesses.
Even disks with very small metallicity can easily contain the mass of
planetesimals required to produce this dust.

A lower mass of solid material may even assist in dust production.
Lower surface density disks contain less material for the largest
growing bodies to accumulate.  
The amount of material that a solid core can directly sweep up
(its isolation mass) increases as surface density to the 3/2 power
\citep[e.g.][]{pollack96}, such that disks with lower surface density
tend to produce a larger quantity of smaller protoplanetary
cores, rather than a few large planets.
In this scenario, high metallicity would translate to larger planets
and a cleaner, less dusty central disk.
An outer fringe of smaller planetesimals, as in the solar system,
could still form at or be scattered to the outer disk edge.

Another possibility is that there is an initial correlation between dust
production and metallicity around young stars, but that this relationship
disappears as the stars age.
\citet{dominik03} find that theoretical models of debris disk
evolution tend to evolve toward the same final dust distributions
over long enough time scales.
While the more sparse disks (0.1 $\MEarth$) evolve on Gyr time
scales, the brightest disks generally decay relatively quickly.
Disk models with initial masses ranging from 1 to 100 $\MEarth$
converge toward the same asymptotic trend in less than a billion
years, such that any initial differences in disk mass become 
unimportant for old systems.
Star-to-star variability in dust emission
may be strongly related to stochastic collisional events,
rather than a simple function of initial disk mass.

\subsection{Spectral Type}\label{sptypsec}

Within our observed range of spectral types 
($\Tstar \simeq 4500-6500$ K), 
we have not found any evidence for a correlation with excess emission. 
The average spectral type is G3 for both the stars with IR
excess and for those without.
The meaning of this flat trend is somewhat ambiguous based on our
limited knowledge of the location of the dust, as well as the limited
range of spectral types in our sample.

\section{Frequency of IR Excess around Solar-Type Stars}\label{frequency}

The preliminary results of our survey contain enough
excess detections at \70um to consider the overall statistics for
emission by cold ($\sim$50 K) dust.
Unlike previous investigations, we achieve photospheric detections
at \70um for most of our sample, and the level of detectable disk
brightness usually extends below $\ld \sim 10^{-5}$.  
More importantly, our selection criteria produce an unbiased sample of 
observations that, combined with accurate knowledge of all
measurement uncertainties, allow for a straightforward determination
not only of the overall frequency of IR excess, but of the
distribution of dust luminosities. 
Our sample is similar to a volume limited survey, 
rather than all-sky IRAS observations, which tend to
pick out distant objects with strong excesses.
Unlike a strict volume limited survey, however, we have maximized our
detection efficiency by concentrating on the targets most likely to
produce high signal-to-noise results. 

The detection rate of IR excess depends both on the stellar emission and
the achievable detection limits.
From Eq.~\ref{ldeq}, the detection limit for each star is
\begin{equation}\label{fluxlev}
\frac{L_{\rm dust}}{L_{\star}}({\rm detectable}) 
= 10^{-5} 
\; \left(\frac{5600 \; {\rm K}}{T_{\star}}\right)^3
\; \frac{3 N_{70} }
{F_{70, \star}}
\end{equation}
where $N_{70}$ is the 1-$\sigma$ error in the flux measurement
(listed in Table~\ref{resultstable}).

\begin{figure}
\begin{center}
\includegraphics[width=4.7in,angle=-90]{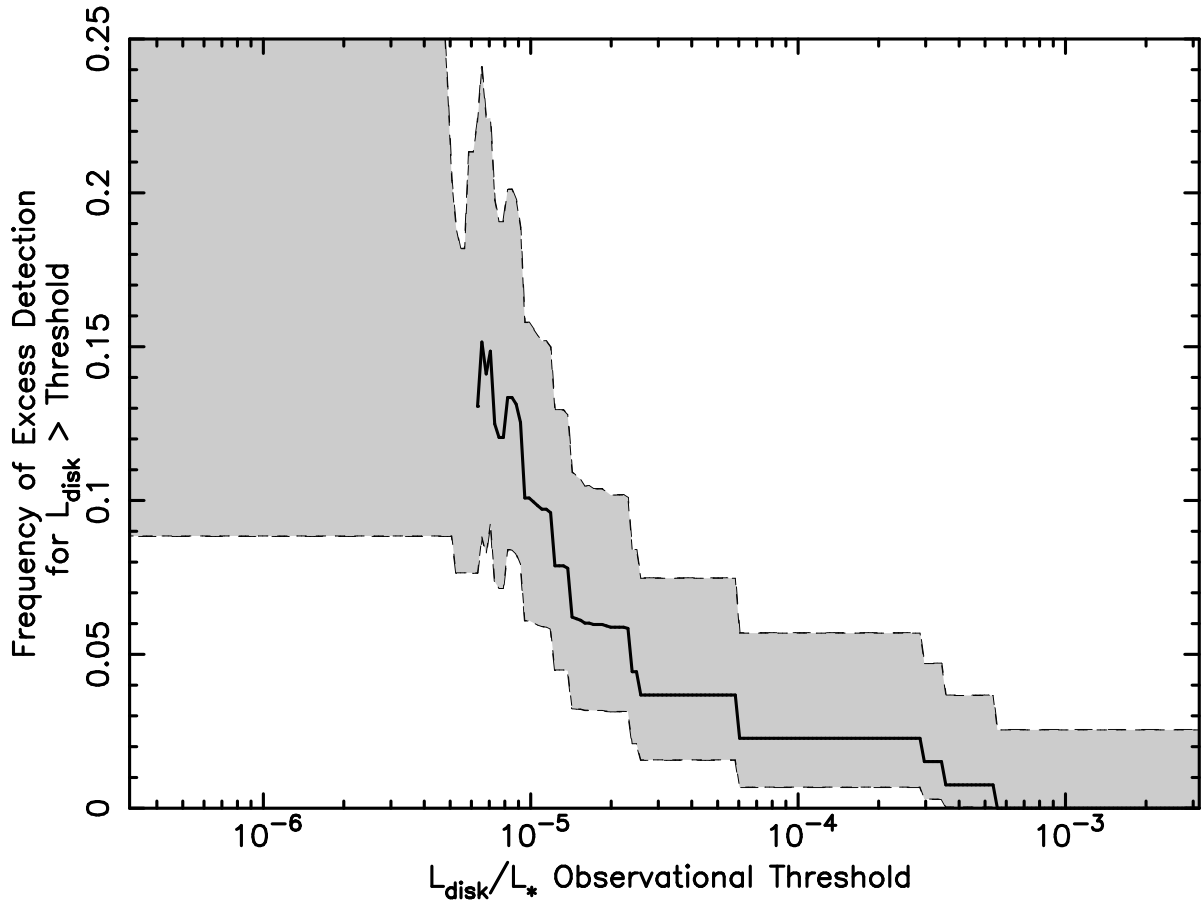} 
\end{center}
\caption{
The frequency of \70um excess detection as a function of
the observational threshold in terms of $\ld$.  
In addition to the 69 observed stars,
a few stars reserved by other programs (Table~\ref{legacytable})
have been included in the figure with appropriate weighting.  
Even including these bright IRAS sources, above 
$\ld \simeq 10^{-3}$ there are no detections.  
The gray region indicates the 1-$\sigma$ limits to the distribution
based on proper binomial statistics.  
The lack of sensitivity below $5 \times 10^{-6}$ is reflected in the
large unconstrained gray region filling the upper-left section of the plot.}
\label{freqvls}
\end{figure}

\begin{figure}
\begin{center}
\includegraphics[width=4.2in]{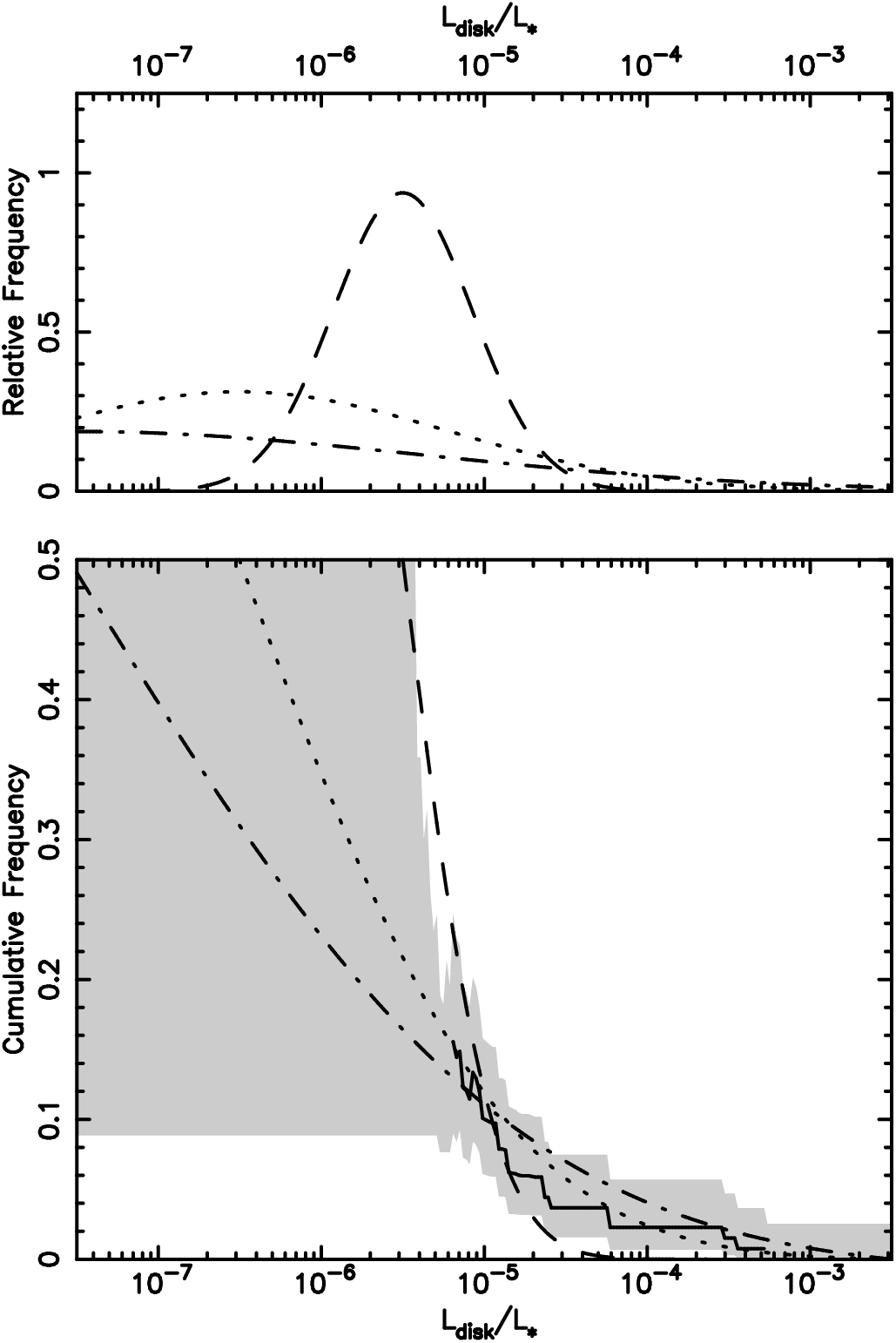} 
\end{center}
\caption{
Disk detection frequency compared with theoretical debris
disk distributions.  Three possibilities are considered: 1) all stars
have disks, with the solar system's level of
emission ($\sim10^{-6.5}$) as average ({\it dotted}), 
2) all stars have disks with average 10 times solar ({\it dashed}),
and 
3) all stars have disks with average 10 times less than solar
({\it dot-dash}).
The relative frequencies of disk luminosity (upper panel) 
are assumed to follow Gaussian distributions,
with the cumulative frequency of disks with $\ld > 10^{-5}$ fixed at
12\% in each case.
The corresponding cumulative frequency distributions are shown in the
lower panel.
The detection frequency within our data is plotted as a solid line for comparison.
As in Fig.~\ref{freqvls}, the region constrained by our observations
is shown in gray (1-$\sigma$).
Of the three curves, the distribution with solar as average (dotted)
is the best fit to the data.
} 
\label{freqthe}
\end{figure}

Fig.~\ref{freqvls} shows how frequently we detect
debris disks above a range of detection 
thresholds.\footnote{The distribution plotted in Fig.~\ref{freqvls} is similar
to, but different from, a true  
cumulative frequency distribution, which always increases monotonically. 
Also note that the standard definition of $\sigma$, the measurement
uncertainty, applies to a Gaussian distribution.
For the binomial distributions considered here, we define the
1-$\sigma$ errors as having the same likelihood as for a Gaussian
distribution, i.e.\ there is a 68\% probability that the true value
lies within the gray region in Figs.~\ref{freqvls}-\ref{freqthe}.}
For each observational threshold (each $\ld$), 
only stars with a lower detectability limit
(Eq.~\ref{fluxlev}) are considered.
The \70um observations are generally very sensitive to
disks with $\ld > 10^{-5}$, with many cleaner fields sensitive to as
low as $\sim 5 \! \times \! 10^{-6}$.   
Below this level, we have no direct measurements of the disk frequency, 
and the 1-$\sigma$ constraints on the frequency of excess
detection (the shaded region in the figure) are not well defined.  

As discussed in \S\ref{sample}, 131 stars meet our selection criteria
for an unbiased sample.  However, four of these are well-known
IR-excess sources that have been reserved by other programs 
(Table~\ref{legacytable}).
To avoid a bias against strong excesses, these stars have also 
been included into the overall statistics of 
Fig.~\ref{freqvls}, with weighting appropriate for the fraction of 
stars currently observed (69/127).

Even with the inclusion of these bright disks,
we find that the frequency of disk detection increases steeply as
the detection limit is extended down to dimmer disks.
While debris disks with $\ld \sim 10^{-3}$ are rare around old solar-type
stars, the disk frequency increases from 2$\,\pm\,$2\% for
disks with $\ld > 10^{-4}$ to 12$\,\pm\,$5\% for $\ld > 10^{-5}$.
Our overall detection rate is in good agreement with the results
of \citet{kim05}, who find five \70um excesses in a sample of 35
solar-type stars, a detection rate of $14 \pm 6\%$.  

With these data, we can start to place the
dusty debris in the solar system into context relative to other solar-type
stars.  Extrasolar planetary systems with architectures very
different from our own continue to be discovered.  With highly
eccentric planets, short-period planets, and resonantly locked planets
all commonly seen around other stars, the solar system may not be a
typical planetary system, nor may its interplanetary dust be typical.
 
Fig.~\ref{freqthe} shows our observed \70um disk brightness distribution
compared with several simple theoretical distributions. 
Three possibilities for the median disk luminosity are considered:
equal to the Kuiper Belt's level of emission ($\sim$$10^{-6.5}$, 
{\it dotted}), ten times above this level ({\it dashed}), 
and ten times below ({\it dot-dash}). 
In each case, we set the frequency of disks with $\ld\ \gapp 10^{-5}$
at 12\%, in accordance with Fig.~\ref{freqvls}.
Also, we assume Gaussian distributions of disk luminosities
(in logarithmic space), resulting in standard deviations of 1.8, 3.0,
and 0.6 decades for the three curves. 
The true distribution of disk luminosities is not a strict Gaussian,
but more likely has an extended tail of strong emitters 
resulting from recent collisional events.
Nonetheless, under the rough assumption that the distribution of
debris disk luminosities follows a Gaussian-shaped profile, our
existing dataset can already limit the theoretical disk distributions
to some extent.  In particular, the possibility that most stars have
disks much brighter than the solar system's ({\it dashed lines})
appears to be inconsistent with the constraints provided by our
observations ({\it gray region}).

\section{Summary}\label{conclusions}

We have searched for circumstellar dust around an unbiased sampling of
69 F5-K5 stars by means of photometric measurements at \24um  and \70um. 
We detected all the stars at \24um with high S/N and 80\% of the stars at
\70um with S/N $> 3$.
Uncertainties in the \spit\ calibration and in the extrapolation of
stellar photospheres to far-IR wavelengths limit our ability to detect
IR excesses with 3-$\sigma$ confidence to $\sim$20\% and $\sim$50\% of
the photospheric levels at 24 and \70um, respectively.  

Of the 69 stars, we have a single detection of excess at \24um, for an
overall detection rate of $\sim$1\%. 
At \70um, seven stars show significant excesses ($>$3-$\sigma$).
When we correct the detection statistics for large-excess stars
intentionally left out of the sample, the incidence of \70um excesses in
this type of star is $13 \pm 5$\%.
With only a single wavelength of excess measurement, the dust
properties for these stars are not well constrained, but are generally  
consistent with Kuiper Belt configurations --
distances from the star of several tens of AU and  temperatures of
$\sim$50 K.  The observed dust luminosities, however, are much
brighter than in the solar system, generally exceeding the Kuiper
Belt's $\ld$ by factors of $\sim$100.   

Cross-correlating the detections of IR excess with stellar parameters
we find no significant correlations in the incidence
of excesses with 1) age, 2) metallicity, or 3) spectral type. 
The restricted range of the sample in age and spectral type may
hide more global correlations that can be explored with broader samples. 
The lack of correlation with metallicity contrasts with 
the known correlation between planet detections
and stellar metallicity, and the expectation that 
higher metal content might result in a
greater number of dust-producing planetesimals.

We have a large enough sample of excess detections 
at \70um to fit the cumulative distribution,
which rises from $\sim$2\% for $\ld > 10^{-4}$
to $\sim$12\% for $\ld > 10^{-5}$.
Under the assumption that the distribution of disk luminosities follows
a Gaussian distribution, the current observations suggest that the
infrared emission  by dust in the Kuiper Belt must be within a factor
of 10, greater or less, of the typical level for an average solar-type star.

While only one star has detectable excess emission at \24um 
\citep[HD69830; see][]{Beichman05comet}, in some
ways we are less sensitive to dust at that wavelength.
Although better instrumentation gives us better sensitivity at \24um
in terms of the relative flux ($F_{\rm dust}/F_{\star}$),
as far as fractional disk luminosity we are only sensitive to disks
with $L_{\rm dust} \gapp 5 \! \times \! 10^{-5} \, L_{\star}$ at \24um, 
an order of magnitude worse than at \70um.
This detection threshold is many orders of magnitude above
the luminosity of the asteroid belt
\citep[$\ld \simeq 10^{-8}$-$10^{-7}$;][]{Dermott02}.  
The disks that we are detecting have typical \70um luminosities around
100 times that of the Kuiper Belt. 
If they also have inner asteroid belts 100 times brighter than our
own, we would still not be able to detect the warm inner dust.
In other words, the observed \70um excess systems could all be
scaled-up replicas of the solar system's dust disk architecture,
differing only in overall magnitude.
These systems could have planets, asteroids, and Kuiper Belt Objects as in our own system,
but simply with a temporarily greater amount of dust due to a recent
collisional event.

\acknowledgments {
This publication makes use of data products from the Two-Micron All
Sky Survey (2MASS), as well as from IPAC/IRSKY/IBIS, SIMBAD, VIZIER, and
the ROE Debris Disks Database website.
B.~Heyburn and S.~Sarkissian contributed to the compilation of data
from these sites.
We would like to thank K.~Grogan, C.~Dominik, G.~Laughlin, and S.~Fajardo-Acosta
for helpful discussions, and an anonymous referee for a careful
reading of the manuscript.

The {\it Spitzer Space Telescope} is operated by the Jet Propulsion
Laboratory, California Institute of Technology, under NASA contract 1407. 
Development of MIPS was funded by NASA through the Jet Propulsion
Laboratory, subcontract 960785.  Some of the research described in
this publication was carried out at the Jet Propulsion Laboratory,
California Institute of Technology, under a contract with the National
Aeronautics and Space Administration.  

Finally, we note that much of the preparation for the observations
described here was carried out by Elizabeth Holmes, who passed away  
in March, 2004.  This work is dedicated to her memory.}

\appendix
\section{Modeling the Stellar Photosphere}
\label{kuruczsec}

Developing accurate spectral models for the photospheres of our target
stars is critical for determining the presence and strength of any IR
excess.  This is particularly true for measurements with low
background noise (i.e.\ \24um) where inaccuracy in our photospheric
models is likely to be the greatest source 
of uncertainty in identifying excess emission.
Accordingly, we have compiled the best available photometric
measurements for our target stars and used this data to extrapolate
from visible/near-IR wavelengths out to 24 and \70um.
Fortunately, the FGK sample is made up of bright, well-known stars
of solar-like spectral types, making the photospheric modeling
relatively straightforward.

From the literature we have assembled visible
photometry in five bands: $U$, $B$, $V$, $R$, and $I$. 
Whenever possible, we derived B and V data from the Hipparcos
satellite measurements \citep{hipparcos} transformed to a common
Johnson color system. 
These Hipparchos magnitudes are typically accurate to $\sim$0.01 mag. 
Data at $U$, $R$, and $I$ come from a wide variety of references,
including compilations by \citet{johnson1975}, \citet{morel1978},
\citet{bessel1990}, \citet{guarinos1992}, \citet{degeus1990}, and
\citet{bessel1990}.
Five near-IR bands ($J$, $H$, $K$/$K_s$, $L$/$L^\prime$, and
$M$) are considered;
data in these bands come from the visible photometry references,
from data compiled in \citet{gezari1993,gezari1999}, 
and from the 2MASS catalog.
For stars with high quality detections,
IRAS measurements at 12 and 25 \um\ are also included;
fluxes from the IRAS Faint Source Catalog \citep{moshir1990} 
have been color corrected based on the stellar effective temperature. 
For most of our sources, 2MASS photometry sets a limiting accuracy 
of $\sim$2\% in our extrapolation to MIPS wavelengths.  
Many stars, however, are bright 
enough ($K_s \lapp 4$) to saturate one or all of the 2MASS bands.
2MASS accuracy in these cases is only 0.10-0.25 mag, such that the
Hipparcos visible photometry plays a greater role in the overall fit.

The compiled data is fit with Kurucz stellar atmosphere models
\citep{kurucz1992,lejeune1997,Castelli03,Kurucz03}, which are
appropriate for the F-K type stars considered here.
Each Kurucz model was integrated over representative filter and
atmospheric passbands, incorporating the effects of spectral lines
that are particularly important in the $U$, $B$, and $V$ bands. 
The Johnson system flux zero points are taken from
\citet{campins1985} and \citet{rieke1985}. 
Flux uncertainties for each photometry band are taken as their
published errors, 
but with an imposed minimum fractional uncertainty of 2\%. 

In addition to the photometric fluxes, each star's observed spectral
type and metallicity ($[Fe/H]$) are given  
as inputs to the program, with assumed errors of
250 K for $T_{eff}$ and 0.25 dex for $[Fe/H]$.
The fitting program steps through a discrete grid of 
effective temperatures spaced every 250 K 
and [Fe/H] values of -1.0, -0.5, -0.2, 0.0, 0.2, 0.5, and 1.0. 
While the logarithm of the stellar surface gravity, $\log{g}$, is
known to vary from 4.32 to 4.60 for F5 to K5 stars \citep{gray1992},
we assume $\log{g}$ = 4.5 for all cases.  
A microturbulent velocity of 2.00 km~s$^{-1}$ is also assumed.

Given the uncertainties for each variable, a minimum-$\chi^2$
fit is obtained, scaling the data to Kurucz-Lejeune models. 
For each wavelength, the r.m.s.\ dispersion in the fits
is very similar to the input uncertainties, as desired.
While the average observed fluxes at each wavelength 
($1/\sigma^2$-weighted, with rejection of 2-$\sigma$ outliers)
are typically within a few percent of the model values,
some bands stand out above this typical $\sim$2\% offset.
At $U$, for example, the data consistently lie an average of 4.7\% below the models. 
Given the difficulty both in calibrating $U$ band photometry 
and in computing $U$ band model photospheres,
the large errors at this wavelength are not unexpected.
There is also a fitting offset at 25 \um, where the IRAS data
sit 4.7\% above the models, an apparent excess that has been
attributed by a number of authors to a small miscalibration of the
IRAS data at 25 \um\ \citep[e.g.][]{cohen1999}. 
With the exception of $U$ band and 25 \um\ data,
the reasonableness of the fits is good within the prescribed errors.
The average offset, combining all wavelengths, is just -0.2\%.

\begin{deluxetable}{l|cc|cccc|cccc} 
\tabletypesize{\scriptsize}
\setlength{\tabcolsep}{0.01in}
\rotate
\tablecaption{FGK Survey Stars
\label{basictable}}
\tablehead{
Star	&	Spectral	&		V &	
	&	Age (Gyr)	&		&
	&	& [Fe/H]	&	&  \\
	&	 Type	 	&	(mag) 	&
Wr/Average$^{c}$  	&	Min  	&
Max  	& Refs 	&	Average	&	Dispersion	
&	\# est.	&  Refs }
\startdata
HD 166$^{a}$	&	K0 Ve	&	6.16	&	5.0	&	0.04	&	5.0	&	Wr,Ba	&	-0.05	&	0.15	&	5	&	Ca,E,M,Hy,Ga	\\
HD 1237$^{a,b}$	&	G6 V	&	6.67	&	2.8	&	0.6	&	2.8	&	Wr,L	&	0.1	&	0.08	&	5	&	Ca,B,Hy,Go,Ga	\\
HD 1581$^{a}$	&	F9 V	&	4.29	&	3.0	&	3.0	&	10.7	&	Wr,L	&	-0.23	&	0.1	&	8	&	T,Ca,E,M,Hy,L	\\
HD 4628$^{a}$	&	K2 V	&	5.85	&	8.1	&	5.5	&	11.0	&	L	&	-0.22	&	0.11	&	6	&	E,Ca,M,Ce,L	\\
HD 7570$^{a}$	&	F8 V	&	5.03	&	4.3	&	1.1	&	7.8	&	M,F,L	&	0.04	&	0.1	&	5	&	T,Ca,M,E,L	\\
HD 10800$^{a}$	&	G2 V	&	5.95	&	7.4	&	-	&	-	&	F	&	-0.03	&	0.08	&	2	&	M	\\
HD 13445$^{a,b}$	&	K1 V	&	6.21	&	5.6	&	3.7	&	7.6	&	L	&	-0.19	&	0.04	&	5	&	Ca,E,B,M,Hy	\\
HD 17051$^{a,b}$	&	G0 V	&	5.46	&	2.4	&	0.5	&	5.1	&	M,B,L,Lw	&	0.09	&	0.11	&	6	&	Ca,M,E,B,Gi,L	\\
HD 20766$^{a}$	&	G2 V	&	5.57	&	5.2	&	2.9	&	7.9	&	L	&	-0.2	&	0.07	&	7	&	T,Ca,E,Hy,L	\\
HD 33262$^{a}$	&	F7 V	&	4.77	&	3.5	&	1.2	&	6.5	&	L	&	-0.21	&	0.07	&	5	&	T,Ca,M,E,L	\\
HD 34411$^{a}$	&	G1.5IV-V	&	4.76	&	6.8	&	4.6	&	9.4	&	Wr,M,Ba,L	&	0.05	&	0.07	&	6	&	T,Ca,M,L,Bo,B	\\
HD 35296$^{a}$	&	F8 Ve	&	5.06	&	3.8	&	0.02	&	7.5	&	C,Ba	&	-0.06	&	0.09	&	4	&	T,Ca,M,C	\\
HD 37394$^{a}$	&	K1 Ve	&	6.30	&	0.5	&	0.3	&	0.9	&	L	&	-0.07	&	0.1	&	7	&	T,Ca,E,Hy,L,Le,Ga	\\
HD 39091$^{a,b}$	&	G1 V	&	5.72	&	5.6	&	3.9	&	7.3	&	M,F	&	0.11	&	0.11	&	5	&	T,M,E,Ca,Hy	\\
HD 43162$^{a}$	&	G5 V	&	6.45	&	0.4	&	-	&	-	&	Wr	&	-0.14	&	0.04	&	2	&	Hy,Ga	\\
HD 43834$^{a}$	&	G6 V	&	5.15	&	7.6	&	5.1	&	10.5	&	L	&	0.04	&	0.12	&	7	&	T,Ca,E,M,Hy,L	\\
HD 50692$^{a}$	&	G0 V	&	5.82	&	4.5	&	-	&	-	&	Wr	&	-0.19	&	0.11	&	2	&	M,E	\\
HD 52711$^{a}$	&	G4 V	&	6.00	&	4.8	&	4.8	&	6.4	&	Wr,Ba	&	-0.16	&	0.03	&	4	&	T,Ca,M,E	\\
HD 55575$^{a}$	&	G0 V	&	5.61	&	4.6	&	4.6	&	10.6	&	Wr,C	&	-0.3	&	0.08	&	6	&	T,Ca,M,C,E,Bo	\\
HD 58855$^{a}$	&	F6 V	&	5.41	&	3.6	&	-	&	-	&	C	&	-0.27	&	0.08	&	5	&	T,Ca,M,C,Ms	\\
HD 62613$^{a}$	&	G8 V	&	6.63	&	3.1	&	3.1	&	5.2	&	Wr,Ba	&	-0.17	&	0.04	&	2	&	E,Hy	\\
HD 68456$^{a}$	&	F5 V	&	4.80	&	2.4	&	-	&	-	&	M	&	-0.28	&	0.07	&	4	&	T,Ca,M,E	\\
HD 69830$^{a}$	&	K0 V	&	6.04	&	4.7	&	0.6	&	4.7	&	Wr,So	&	0.0	&	0.06	&	4	&	Ca,E,Hy,M	\\
HD 71148$^{a}$	&	G5 V	&	6.39	&	4.7	&	4.7	&	5.6	&	Wr,Ba	&	-0.05	&	0.18	&	3	&	M,E,Hy	\\
HD 72905$^{a}$	&	G1.5 V	&	5.71	&	0.4	&	0.01	&	0.4	&	Wr,Ba,W	&	-0.04	&	0.1	&	5	&	T,Ca,M,E,Ga	\\
HD 75732$^{a,b}$	&	G8 V	&	6.04	&	6.5	&	3.6	&	6.5	&	Wr,B,Lw	&	0.31	&	0.13	&	7	&	T,Ca,B,M,Hy,Ce	\\
HD 76151$^{a}$	&	G3 V	&	6.08	&	1.8	&	0.8	&	4.1	&	M,L	&	0.09	&	0.05	&	5	&	T,Ca,M,E,L	\\
HD 84117$^{a}$	&	G0 V	&	4.98	&	4.2	&	2.5	&	6.0	&	M,F	&	-0.15	&	0.07	&	4	&	T,M,E,Hy	\\
HD 84737$^{a}$	&	G0.5 Va	&	5.16	&	11.7	&	4.3	&	11.7	&	Wr,M,F,Ba,L	&	0.04	&	0.04	&	5	&	T,Ca,M,E,L	\\
HD 88230$^{a}$	&	K2 Ve	&	6.75	&	4.7	&	-	&	-	&	L	&	-0.02	&	0.61	&	4	&	Ca,Ce,B,Ma	\\
HD 90839$^{a}$	&	F8 V	&	4.88	&	3.4	&	0.2	&	5.2	&	Wr,C,Ba,L	&	-0.15	&	0.07	&	5	&	T,Ca,M,C,L	\\
HD 95128$^{a,b}$	&	G0 V	&	5.1	&	6.0	&	3.9	&	8.3	&	Wr,B,Ba,L,Lw	&	-0.01	&	0.05	&	7	&	T,Ca,M,C,Gi,L,Lw	\\
HD 101501$^{a}$	&	G8 Ve	&	5.39	&	1.1	&	0.5	&	2.5	&	Wr,Ba,L	&	-0.2	&	0.21	&	8	&	T,Ca,E,M,Ce,Hy,L	\\
HD 102870$^{a}$	&	F9 V	&	3.66	&	4.5	&	2.2	&	7.5	&	Wr,M,Ba,L	&	0.14	&	0.06	&	7	&	T,Ca,M,C,L,Bo,B	\\
HD 114710$^{a}$	&	F9.5 V	&	4.31	&	2.3	&	1.8	&	9.6	&	Wr,C,Ba,L	&	0.03	&	0.07	&	8	&	T,Ca,M,C,Ms,L,Le,Bo	\\
HD 115383$^{a}$	&	G0 V	&	5.26	&	0.4	&	0.1	&	5.7	&	Wr,M,C,Ba,L	&	0.02	&	0.06	&	8	&	Hy,M,C,Ca,L	\\
HD 115617$^{a}$	&	G5 V	&	4.81	&	6.3	&	4.4	&	9.6	&	Wr,L	&	0.01	&	0.04	&	5	&	T,Ca,E,L,Bo	\\
HD 117043$^{a}$	&	G6	&	6.58	&	-	&	-	&	-	&	-	&	0.22	&	-	&	1	&	Hy	\\
HD 117176$^{a,b}$	&	G2.5 Va	&	5.05	&	5.4	&	5.4	&	12.1	&	Wr,B,Ba,L,Lw	&	-0.06	&	0.04	&	9	&	T,Ca,B,M,Ce,Ms,Gi,Lw	\\
HD 122862$^{a}$	&	G2.5 IV	&	6.08	&	6.1	&	4.8	&	7.4	&	M,F	&	-0.16	&	0.05	&	3	&	M,F,E	\\
HD 126660$^{a}$	&	F7 V	&	4.10	&	2.8	&	0.4	&	4.7	&	M,Ba,L	&	-0.11	&	0.11	&	6	&	T,Ca,M,Ce,L	\\
HD 130948$^{a}$	&	G2 V	&	5.94	&	0.9	&	0.9	&	10	&	Wr,C,W	&	-0.02	&	0.14	&	6	&	Ca,M,C,E,Ga,B	\\
HD 133002$^{a}$	&	F9 V	&	5.71	&	2.5	&	-	&	-	&	F	&	-	&	-	&	0	&	-	\\
HD 134083$^{a}$	&	F5 V	&	4.98	&	1.7	&	0.9	&	2.5	&	L	&	-0.01	&	0.09	&	5	&	T,Ca,M,E,L	\\
HD 136064$^{a}$	&	F8 V	&	5.21	&	4.6	&	4.5	&	4.8	&	M,F	&	-0.04	&	0.02	&	7	&	T,Ca,M	\\
HD 142373$^{a}$	&	F9 V	&	4.67	&	8.1	&	7.1	&	9.7	&	Wr,M,C,Ba,L	&	-0.45	&	0.07	&	7	&	T,Ca,M,C,Ms,L,Bo	\\
HD 142860$^{a}$	&	F6 IV	&	3.88	&	2.9	&	2.4	&	4.3	&	Wr,C,L	&	-0.16	&	0.08	&	8	&	T,Ca,C,M,L	\\
HD 143761$^{a,b}$	&	G0 V	&	5.47	&	7.4	&	7.4	&	12.1	&	Wr,M,C,B,Ba,Lw	&	-0.23	&	0.05	&	6	&	T,Ca,M,C,Gi,Bo	\\
HD 146233$^{a}$	&	G1 V	&	5.56	&	4.6	&	4.4	&	4.6	&	Wr,M	&	0.06	&	0.04	&	4	&	T,Ca,M,E	\\
HD 149661$^{a}$	&	K2 V	&	5.86	&	1.2	&	1.2	&	3.1	&	Wr,L	&	0.11	&	0.26	&	6	&	E,Ca,M,Hy,Ce,L	\\
HD 152391$^{a}$	&	G8 V	&	6.74	&	0.6	&	0.3	&	0.6	&	Wr,Ba	&	-0.11	&	0.07	&	4	&	E,M,Hy,Ga	\\
HD 157214$^{a}$	&	G2 V	&	5.46	&	6.5	&	5.6	&	12.0	&	Wr,Ba,L	&	-0.36	&	0.04	&	8	&	T,Ca,M,Hy,Ce,Ms,L	\\
HD 166620$^{a}$	&	K2 V	&	6.49	&	5.0	&	4.4	&	12.0	&	Wr,L	&	-0.18	&	0.14	&	7	&	T,E,Ca,Ce,Hy,L,Le	\\
HD 168151$^{a}$	&	F5 V	&	5.04	&	2.5	&	2.4	&	2.7	&	M,C,F	&	-0.28	&	0.08	&	4	&	T,Ca,M,C	\\
HD 173667$^{a}$	&	F6 V	&	4.26	&	3.4	&	2.1	&	3.8	&	Wr,M,L	&	-0.12	&	0.05	&	4	&	T,Ca,M,L	\\
HD 181321$^{a}$	&	G5 V	&	6.55	&	0.5	&	-	&	-	&	W	&	-	&	-	&	0	&	-	\\
HD 185144$^{a}$	&	K0 V	&	4.76	&	3.2	&	3.2	&	6.9	&	Wr,Ba,L	&	-0.23	&	0.13	&	6	&	T,E,Ca,M,L,Le	\\
HD 186408$^{a}$	&	G1.5 Vb	&	5.96	&	10.4
&	-	&	-	&	N	&  0.08	&	0.10	&	3	&	T,Ca	\\
HD 186427$^{a,b}$	&	G3 V	&	6.29	&	8.7	&	8.3	&	9.1	&	B,Lw	&	0.06	&	0.04	&	6	&	T,Ca,E,B,Gi,Bo	\\
HD 188376$^{a}$	&	G5 V	&	4.77	&	7.4	&	-	&	-	&	Wr	&	-0.02	&	0.15	&	2	&	T,Ca	\\
HD 189567$^{a}$	&	G2 V	&	6.15	&	4.5	&	-	&	-	&	Wr	&	-0.26	&	0.07	&	6	&	T,Ca,E,M,Hy	\\
HD 190248$^{a}$	&	G7 IV	&	3.62	&	5.3	&	-	&	-	&	L	&	0.36	&	0.11	&	6	&	T,Ca,E,M,Hy	\\
HD 196378$^{a}$	&	F8 V	&	5.18	&	6	&	5.8	&	6.2	&	M,F	&	-0.39	&	0.06	&	3	&	T,Ca,M	\\
HD 197692$^{a}$	&	F5 V	&	4.19	&	1.9	&	1.4	&	2.5	&	L	&	0.0	&	0.06	&	6	&	T,Ca,M,L	\\
HD 203608$^{a}$	&	F8 V	&	4.28	&	10.2	&	6.5	&	14.5	&	C,F,L	&	-0.65	&	0.11	&	5	&	T,Ca,M,C,L	\\
HD 206860$^{a}$	&	G0 V	&	6.02	&	5.0	&	0.09	&	9.9	&	C,Ba	&	-0.12	&	0.08	&	5	&	Ca,M,C,E,Ga	\\
HD 210277$^{a,b}$	&	G0V	&	6.63	&	6.8	&	6.8	&	6.9	&	Wr,B,Lw,L	&	0.23	&	0.01	&	3	&	Ca,B,Hy	\\
HD 216437$^{a,b}$	&	G2.5IV	&	6.13	&	7.2	&	6.3	&	8.0	&	F,R	&	0.2	&	0.1	&	4	&	M,Ca,R	\\
HD 221420$^{a}$	&	G2 V	&	5.89	&	5.5	&	-	&	-	&	F	&	0.55	&	-	&	1	&	M	\\
\hline
HD 693$^{}$	&	F5 V	&	4.95	&	5.2	&	4.3	&	5.9	&	M,C,L	&	-0.4	&	0.03	&	5	&	T,Ca,M,C,L	\\
HD 3302$^{}$	&	F6 V	&	5.56	&	7.8	&	2.1	&	7.8	&	Wr,M,F	&	-0.23	&	0.29	&	2	&	M,E	\\
HD 3651$^{b}$	&	K0 V	&	5.97	&	5.9	&	-	&	-	&	Wr	&	-0.03	&	0.1	&	5	&	T,Ce,E,Ca,M	\\
HD 3795$^{}$	&	G3 V	&	6.23	&	7.2	&	7.2	&	10.8	&	Wr,F	&	-0.67	&	0.04	&	4	&	Ca,Ms	\\
HD 3823$^{}$	&	G1 V	&	5.96	&	5.5	&	4.9	&	10.5	&	Wr,M,F	&	-0.32	&	0.1	&	4	&	Ca,M,E	\\
HD 4307$^{}$	&	G2 V	&	6.22	&	7.8	&	7.2	&	7.8	&	Wr,M	&	-0.24	&	0.04	&	5	&	T,Ca,M,E,Bo	\\
HD 9826$^{b}$	&	F8 V	&	4.16	&	6.3	&	2.3	&	6.3	&	Wr,M,B,Ba,L,Lw	&	0.04	&	0.07	&	6	&	T,Ca,M,C,Gi,L	\\
HD 10476$^{}$	&	K1 V	&	5.34	&	4.6	&	-	&	-	&	Wr	&	-0.16	&	0.05	&	6	&	T,Ca,E,Ce,M,Le	\\
HD 10697$^{b}$	&	G5 IV	&	6.36	&	7.4	&	7.4	&	7.9	&	Wr,F,Lw	&	0.1	&	0.08	&	5	&	Ca,B,Ms,Go,Gi	\\
HD 13555$^{}$	&	F5 V	&	5.28	&	2.7	&	2.7	&	2.8	&	C,F	&	-0.3	&	0.05	&	4	&	T,Ca,M,C	\\
HD 14412$^{}$	&	G5 V	&	6.42	&	3.3	&	3.3	&	12	&	Wr,L	&	-0.42	&	0.66	&	4	&	Ca,E,Hy,L	\\
HD 14802$^{}$	&	G2 V	&	5.27	&	6.8	&	5.0	&	6.8	&	Wr,M,L	&	-0.09	&	0.04	&	6	&	Ca,M,E,Ce,L	\\
HD 15335$^{}$	&	G0 V	&	5.97	&	7.8	&	6.9	&	8.1	&	Wr,M,C,F	&	-0.21	&	0.04	&	4	&	T,Ca,M,C	\\
HD 15798$^{}$	&	F5 V	&	4.79	&	3.2	&	2.5	&	4.1	&	M,C,F	&	-0.25	&	0.02	&	4	&	T,Ca,M,C	\\
HD 16160$^{}$	&	K3 V	&	5.80	&	-	&	-	&	-	&	-	&	-0.08	&	0.04	&	3	&	E,Hy,Ce	\\
HD 17925$^{}$	&	K1 V	&	6.15	&	0.2	&	0.04	&	0.5	&	L,W	&	-0.02	&	0.11	&	7	&	T,Ca,E,M,Hy,L,Le	\\
HD 19373$^{}$	&	G0 V	&	4.12	&	5.9	&	2.1	&	8.1	&	Wr,C,Ba,L	&	0.09	&	0.08	&	6	&	T,Ca,C,E,L,Bo	\\
HD 20630$^{}$	&	G5 Ve	&	4.92	&	0.3	&	0.2	&	0.4	&	Ba,L	&	0.0	&	0.09	&	8	&	T,Ca,E,M,Mu,Ce,L	\\
HD 20807$^{}$	&	G1 V	&	5.30	&	7.9	&	4.4	&	12	&	L	&	-0.2	&	0.04	&	8	&	T,Ca,E,M,Hy,L	\\
HD 22484$^{}$	&	F8 V	&	4.36	&	8.3	&	4.6	&	8.4	&	Wr,C,M,Ba,L	&	-0.11	&	0.06	&	6	&	T,Ca,M,C,L,Bo	\\
HD 26923$^{}$	&	G0 IV	&	6.38	&	-	&	-	&	-	&	-	&	0.0	&	0.06	&	5	&	Ca,M,R,Ga	\\
HD 30495$^{}$	&	G1 V	&	5.56	&	1.3	&	0.2	&	1.3	&	Wr,L	&	0.0	&	0.08	&	6	&	Ca,M,E,Hy,L,Ga	\\
HD 30652$^{}$	&	F6 V	&	3.24	&	1.5	&	-	&	-	&	Wr	&	0.0	&	0.06	&	6	&	T,M,E,Ca,Ce	\\
HD 33564$^{}$	&	F6 V	&	5.14	&	3.5	&	-	&	-	&	M	&	-0.12	&	0.01	&	2	&	M,E	\\
HD 34721$^{}$	&	G0 V	&	6.02	&	6.2	&	3.8	&	6.2	&	Wr,M	&	-0.13	&	0.09	&	5	&	Ca,M,E,Hy	\\
HD 69897$^{}$	&	F6 V	&	5.18	&	3.5	&	1.9	&	4.7	&	Wr,C,L	&	-0.26	&	0.04	&	5	&	T,Ca,M,C,L	\\
HD 86728$^{}$	&	G3 Va	&	5.45	&	6.9	&	1.5	&	6.9	&	Wr,F	&	0.0	&	-	&	1	&	M	\\
HD 94388$^{}$	&	F6 V	&	5.29	&	3.2	&	-	&	-	&	M	&	0.09	&	0.06	&	3	&	T,Ca,M	\\
HD 102438$^{}$	&	G5 V	&	6.56	&	-	&	-	&	-	&	-	&	-0.21	&	0.26	&	3	&	E,Hy,M	\\
HD 103932$^{}$	&	K5 V	&	7.10	&	-	&	-	&	-	&	-	&	0.16	&	-	&	1	&	Ca	\\
HD 104731$^{}$	&	F6 V	&	5.21	&	1.8	&	1.7	&	2	&	M,F	&	-0.15	&	0.05	&	3	&	Ca,M	\\
HD 110897$^{}$	&	G0 V	&	6.02	&	9.7	&	4.9	&	14.5	&	C,Ca	&	-0.49	&	0.06	&	5	&	T,Ca,M,C,Bo	\\
HD 111395$^{}$	&	G7 V	&	6.37	&	1.2	&	-	&	-	&	Wr	&	0.07	&	0.1	&	3	&	E,M,Hy	\\
HD 112164$^{}$	&	G1 V	&	5.96	&	3.4	&	3.2	&	3.6	&	M,F	&	0.25	&	0.1	&	4	&	T,Ca,M	\\
HD 114613$^{}$	&	G3 V	&	4.93	&	5.3	&	-	&	-	&	F	&	-	&	-	&	0	&	-	\\
HD 118972$^{}$	&	K1	&	7.02	&	-	&	-	&	-	&	-	&	-0.05	&	-	&	1	&	Ga	\\
HD 120690$^{}$	&	G5 V	&	6.52	&	2.2	&	-	&	-	&	Wr	&	-0.1	&	0.02	&	3	&	Ca,E,Hy	\\
HD 127334$^{}$	&	G5 V	&	6.44	&	6.9	&	6.9	&	15.9	&	Wr,C	&	0.11	&	0.04	&	4	&	T,Ca,C,E	\\
HD 131156$^{}$	&	G8 V	&	4.55	&	-	&	-	&	-	&	-	&	-0.01	&	0.23	&	4	&	Ca	\\
HD 154088$^{}$	&	G8IV-V	&	6.68	&	5.9	&	4.4	&	12.0	&	Wr,L	&	0.29	&	0.01	&	2	&	Hy,M	\\
HD 181655$^{}$	&	G8 V	&	6.36	&	4.6	&	4.6	&	11.1	&	Wr,F	&	0.05	&	-	&	1	&	E	\\
HD 190007$^{}$	&	K4 V	&	7.60	&	-	&	-	&	-	&	-	&	-	&	-	&	0	&	-	\\
HD 191408$^{}$	&	K3 V	&	5.41	&	7.9	&	4.4	&	12.0	&	L	&	-0.4	&	0.12	&	6	&	T,Ca,E,Hy,L	\\
HD 193664$^{}$	&	G3 V	&	5.98	&	4.7	&	4.7	&	4.7	&	M,Ba	&	-0.1	&	0.09	&	6	&	T,M,E,Hy,Ca	\\
HD 196761$^{}$	&	G8 V	&	6.44	&	4.3	&	-	&	-	&	Wr	&	-0.43	&	0.24	&	2	&	E,Hy	\\
HD 207129$^{}$	&	G0 V	&	5.64	&	5.8	&	4.3	&	8.3	&	M,L	&	-0.08	&	0.04	&	5	&	Ca,M,E,L	\\
HD 209100$^{}$	&	K5 Ve	&	4.83	&	1.4	&	0.8	&	2.0	&	L	&	0.01	&	0.1	&	3	&	Ca,M,L	\\
HD 210302$^{}$	&	F6 V	&	4.99	&	5.4	&	2.5	&	5.4	&	Wr,M	&	0.05	&	0.11	&	4	&	T,Ca,M,E	\\
HD 210918$^{}$	&	G5 V	&	6.29	&	3.9	&	-	&	-	&	M	&	-0.1	&	0.14	&	6	&	Ca,E,M,Hy	\\
HD 212330$^{}$	&	G3 IV	&	5.40	&	7.9	&	-	&	-	&	R	&	-0.04	&	0.09	&	6	&	T,Ca,Hy,R	\\
HD 216803$^{}$	&	K4 V	&	6.60	&	-	&	-	&	-	&	-	&	0.08	&	0.01	&	2	&	Sa,M	\\
HD 217014$^{b}$	&	G4 V	&	5.52	&	7.4	&	4.4	&	10.0	&	Wr,B,Ba,L,Lw	&	0.17	&	0.03	&	8	&	T,Ca,E,B,Go,Gi,L	\\
HD 217813$^{}$	&	G5	&	6.73	&	0.7	&	0.7	&	5.6	&	Wr,M	&	-0.02	&	0.07	&	4	&	M,Hy,Ga	\\
HD 219134$^{}$	&	K3 V	&	5.67	&	12.6	&	-	&	-	&	L	&	0.05	&	0.14	&	8	&	T,E,Ca,Mu,Hy,Ce,Le,Bo	\\
HD 220182$^{}$	&	K1	&	7.45	&	0.3	&	0.3	&	0.3	&	Wr,Ba	&	-0.05	&	0.11	&	4	&	E,M,Hy,Ga	\\
HD 222143$^{}$	&	G5	&	6.67	&	-	&	-	&	-	&	-	&	0.08	&	-	&	1	&	Hy	\\
HD 222368$^{}$	&	F7 V	&	4.19	&	3.9	&	2.7	&	5.2	&	M,Ba,L	&	-0.15	&	0.05	&	6	&	T,Ca,M,C,L,Bo	\\
HD 225239$^{}$	&	G2 V	&	6.18	&	-	&	-	&	-	&	-	&	-0.47	&	0.04	&	2	&	T,Ca	\\
\enddata 
\tablenotetext{a}{Observed}
\tablenotetext{b}{Known planet-bearing star}
\tablenotetext{c}{Age from \citet{Wright04} or an average
of other estimates if Wright data is unavailable.}
\tablecomments{Spectral types from SIMBAD. Visual magnitudes as quoted
in SIMBAD, typically from the Hipparcos satellite.}
\tablerefs{See Table~\ref{reftable}}
\end{deluxetable} 

\begin{deluxetable}{l|ccc|ccccccc} 
\tabletypesize{\small}
\setlength{\tabcolsep}{0.03in}
\tablecaption{Measured and predicted fluxes at 24 and \70um (in mJy)} 
\tablehead{  & & \24um  & & & & & \70um &  & & 	\\
 HD \#  &  $F_{\rm MIPS}$  &  $F_{\star}$  &  $F_{\rm MIPS}/F_{\star}$  
&  $F_{\rm MIPS}$  &  $F_{\star}$  &  $F_{\rm MIPS}/F_{\star}$ & S/N 
&  $\chi_{70}$ $^{a}$ & $F_{\rm dust}$ $^{b}$ & $\ld$ $^{c}$ }
\startdata
166$^{d}$ & 158.0 & 144.9 & 1.09 & 94.9 $\pm$ 4.0 & 16.3 & 5.8 	& 23.3 & 19.8 & 90.4 & 6.8 $\times 10^{-5}$ \\
1237 & 82.9 & 88.7 & 0.94 & 10.0 $\pm$ 2.9 & 10.1 & 1.0 	& 3.8 & 0.0 & & $<$9.0$\times 10^{-6}$ \\
1581 & 545.9 & 573.0 & 0.95 & 81.0 $\pm$ 12.3 & 64.9 & 1.2 	& 7.5 & 1.3 &   & $<$6.9 $\times 10^{-6}$ \\
4628 & 278.6 & 287.2 & 0.97 & 23.9 $\pm$ 9.2 & 32.7 & 0.7 	& 2.5 & -1.0 &   & $<$1.2 $\times 10^{-5}$ \\
7570 & 255.4 & 241.7 & 1.06 & 41.2 $\pm$ 7.6 & 27.2 & 1.5 	& 5.6 & 1.8 &   & $<$9.9 $\times 10^{-6}$ \\
10800 & 124.7 & 120.5 & 1.03 & 17.1 $\pm$ 4.3 & 13.6 & 1.3 	& 5.1 & 0.8 &   & $<$1.2 $\times 10^{-5}$ \\
13445 & 162.9 & 166.9 & 0.98 & 3.9 $\pm$ 6.3 & 19.0 & 0.2 	& 0.7 & -2.4 & & $<$4.1$\times 10^{-6}$ \\
17051 & 166.8 & 161.7 & 1.03 & 20.1 $\pm$ 4.1 & 18.1 & 1.1 	& 5.3 & 0.5 & & $<$6.7$\times 10^{-6}$ \\
20766 & 189.1 & 201.7 & 0.94 & 25.6 $\pm$ 5.4 & 22.9 & 1.1 	& 5.3 & 0.5 &   & $<$8.2 $\times 10^{-6}$ \\
33262$^{d}$ & 326.3 & 312.0 & 1.05 & 60.6 $\pm$ 7.3&35.4& 1.7 	& 9.0 & 3.5 & 29.0 &6.0 $\times 10^{-6}$ \\
34411 & 365.2 & 362.3 & 1.01 & 31.2 $\pm$ 11.7 & 40.8 & 0.8 	& 2.6 & -0.8 &   & $<$6.2 $\times 10^{-6}$ \\
35296 & 240.4 & 238.8 & 1.01 & 24.1 $\pm$ 8.5 & 27.0 & 0.9 	& 3.1 & -0.3 &   & $<$6.2 $\times 10^{-6}$ \\
37394 & 142.2 & 155.3 & 0.92 & 29.7 $\pm$ 7.6 & 17.6 & 1.7 	& 4.7 & 1.6 &   & $<$4.1 $\times 10^{-5}$ \\
39091 & 139.9 & 150.4 & 0.93 & 21.5 $\pm$ 3.6 & 17.0 & 1.3 	& 6.8 & 1.3 & & $<$9.0$\times 10^{-6}$ \\
43162 & 95.3 & 109.3 & 0.87 & 13.5 $\pm$ 2.9 & 12.4 & 1.1 	& 5.2 & 0.4 &   & $<$7.8 $\times 10^{-6}$ \\
43834 & 312.6 & 290.4 & 1.08 & 39.0 $\pm$ 7.0 & 32.6 & 1.2 	& 5.7 & 0.9 &   & $<$8.9 $\times 10^{-6}$ \\
50692 & 137.2 & 138.9 & 0.99 & 10.7 $\pm$ 5.2 & 15.7 & 0.7 	& 2.0 & -1.0 &   & $<$5.7 $\times 10^{-6}$ \\
52711 & 117.0 & 116.2 & 1.01 & 11.1 $\pm$ 3.7 & 13.1 & 0.8 	& 2.7 & -0.5 &   & $<$7.0 $\times 10^{-6}$ \\
55575 & 169.4 & 167.7 & 1.01 & 27.5 $\pm$ 5.4 & 19.0 & 1.4 	& 5.5 & 1.6 &   & $<$1.1 $\times 10^{-5}$ \\
58855 & 154.4 & 149.5 & 1.03 & 14.6 $\pm$ 4.3 & 17.0 & 0.9 	& 4.1 & -0.5 &   & $<$4.6 $\times 10^{-6}$ \\
62613 & 83.7 & 91.8 & 0.91 & 10.1 $\pm$ 2.9 & 10.4 & 1.0 	& 3.6 & -0.1 &   & $<$8.6 $\times 10^{-6}$ \\
68456 & 257.5 & 223.6 & 1.15 & 31.5 $\pm$ 7.8 & 25.4 & 1.2 	& 5.3 & 0.8 &   & $<$7.3 $\times 10^{-6}$ \\
69830$^{e}$ & 230.4 & 158.5& 1.45 &19.3 $\pm$ 4.0&17.9& 1.1 	& 4.9 & 0.4 &   & $<$9.2 $\times 10^{-6}$ \\
71148 & 82.4 & 81.3 & 1.01 & 6.1 $\pm$ 2.6 & 9.2 & 0.7 		& 2.4 & -1.2 &   & $<$5.1 $\times 10^{-6}$ \\
72905$^{d}$ & 165.2 & 154.1 & 1.07 & 41.4 $\pm$ 4.1 &17.4& 2.4	& 11.4 & 5.9 & 27.7 & 1.6 $\times 10^{-5}$ \\
75732 & 172.8 & 162.7 & 1.06 & 18.9 $\pm$ 4.5 & 18.2 & 1.0 	& 4.4 & 0.2 & & $<$8.3$\times 10^{-6}$ \\
76151$^{d}$ & 124.5 & 123.4 & 1.01 & 30.5 $\pm$ 3.9 & 13.9&2.2	& 8.3 & 4.2 & 19.1 & 1.4 $\times 10^{-5}$ \\
84117 & 245.9 & 255.5 & 0.96 & 22.5 $\pm$ 19.3 & 29.0 & 0.8 	& 1.5 & -0.3 &   & $<$1.5 $\times 10^{-5}$ \\
84737 & 252.4 & 253.3 & 1.00 & 30.2 $\pm$ 6.7 & 28.5 & 1.1 	& 5.3 & 0.3 &   & $<$6.4 $\times 10^{-6}$ \\
88230 & 432.7 & 456.9 & 0.95 & 36.2 $\pm$ 8.7 & 52.6 & 0.7 	& 4.4 & -1.9 &   & $<$3.7 $\times 10^{-6}$ \\
90839 & 277.9 & 282.1 & 0.98 & 30.3 $\pm$ 6.4 & 31.9 & 1.0 	& 5.0 & -0.2 &   & $<$4.1 $\times 10^{-6}$ \\
95128 & 259.5 & 265.9 & 0.98 & 29.1 $\pm$ 6.0 & 30 & 1.0 	& 5.5 & -0.1 & & $<$4.8$\times 10^{-6}$ \\
101501 & 262.0 & 288.1 & 0.91 & 30.6 $\pm$ 6.9 & 32.6 & 0.9 	& 5.1 & -0.3 &   & $<$6.1 $\times 10^{-6}$ \\
102870 & 887.6 & 856.3 & 1.04 & 124.1 $\pm$ 18.0 & 96.5 & 1.3 	& 7.2 & 1.5 &   & $<$7.2 $\times 10^{-6}$ \\
114710 & 509.5 & 544.2 & 0.94 & 45.6 $\pm$ 10.3 & 61.5 & 0.7 	& 4.5 & -1.5 &   & $<$2.0 $\times 10^{-6}$ \\
115383 & 219.3 & 229.5 & 0.96 & 16.5 $\pm$ 5.3 & 25.9 & 0.6 	& 2.7 & -1.8 &   & $<$2.1 $\times 10^{-6}$ \\
115617$^{d}$ & 451.1 & 491.0 & 0.92 & 185.6 $\pm$ 16.6&55.7&3.3 & 16.1 & 7.8 & 149.3 & 2.7 $\times 10^{-5}$ \\
117043 & 86.3 & 77.4 & 1.11 & 17 $\pm$ 3.5 & 8.7 & 2.0 		& 5.5 & 2.4 &   & $<$2.3 $\times 10^{-5}$ \\
117176$^{d}$ & 373.6 & 395 & 0.95 & 77.4 $\pm$ 10.2 & 44.8 & 1.7 	& 8.8 &  3.2 & 37.6 & 1.0$\times 10^{-5}$ \\ 
122862 & 105.6 & 108.2 & 0.98 & 14.3 $\pm$ 3.4 & 12.2 & 1.2 	& 4.2 & 0.6 &   & $<$1.0 $\times 10^{-5}$ \\
126660 & 560.0 & 574.6 & 0.97 & 61.6 $\pm$ 10.7 & 65.1 & 0.9 	& 5.9 & -0.3 &   & $<$3.2 $\times 10^{-6}$ \\
130948 & 117.2 & 123.6 & 0.95 & 7.3 $\pm$ 3.3 & 14.0 & 0.5 	& 2.3 & -2 &   & $<$2.4 $\times 10^{-6}$ \\
133002 & 202.2 & 219.0 & 0.92 & 21.4 $\pm$ 4.5 & 25.0 & 0.9 	& 5.2 & -0.8 &   & $<$3.3 $\times 10^{-6}$ \\
134083 & 203.5 & 218.3 & 0.93 & 30.2 $\pm$ 6.5 & 24.8 & 1.2 	& 4.8 & 0.8 &   & $<$6.3 $\times 10^{-6}$ \\
136064 & 208.4 & 206.2 & 1.01 & 18.4 $\pm$ 5.4 & 23.3 & 0.8 	& 3.2 & -0.9 &   & $<$3.6 $\times 10^{-6}$ \\
142373 & 421.5 & 407.1 & 1.04 & 29.7 $\pm$ 9.4 & 46.1 & 0.6 	& 3.4 & -1.8 &   & $<$2.1 $\times 10^{-6}$ \\
142860 & 647.5 & 703.6 & 0.92 & 61.2 $\pm$ 14.4 & 79.8 & 0.8 	& 4.5 & -1.3 &   & $<$2.3 $\times 10^{-6}$ \\
143761 & 201.8 & 192.5 & 1.05 & 27.8 $\pm$ 6.1 & 21.7 & 1.3 	& 5.0 & 1.0 &   & $<$9.6$\times 10^{-6}$ \\
146233 & 183.3 & 172.3 & 1.06 & 20.3 $\pm$ 6.8 & 19.3 & 1.1 	& 3.0 & 0.2 &   & $<$8.2 $\times 10^{-6}$ \\
149661 & 213.4 & 229.9 & 0.93 & 30.4 $\pm$ 7.3 & 26.1 & 1.2 	& 4.4 & 0.6 &   & $<$2.1 $\times 10^{-5}$ \\
152391 & 83.3 & 84.4 & 0.99 & 11.7 $\pm$ 3.6 & 9.5 & 1.2 	& 3.5 & 0.6 &   & $<$1.4 $\times 10^{-5}$ \\
157214 & 217.1 & 225.6 & 0.96 & 23.6 $\pm$ 5.2 & 25.6 & 0.9 	& 4.5 & -0.4 &   & $<$5.3 $\times 10^{-6}$ \\
166620 & 146.7 & 160.4 & 0.91 & 5.9 $\pm$ 5.3 & 18.3 & 0.3 	& 1.3 & -2.3 &   & $<$3.9 $\times 10^{-6}$ \\
168151 & 209.2 & 210.7 & 0.99 & 21.3 $\pm$ 4.8 & 23.9 & 0.9 	& 5.0 & -0.5 &   & $<$3.1 $\times 10^{-6}$ \\
173667 & 445.3 & 427.3 & 1.04 & 68.7 $\pm$ 11.7 & 48.3 & 1.4 	& 6.6 & 1.8 &   & $<$8.5 $\times 10^{-6}$ \\
181321 & 80.9 & 80.5 & 1.01 & 3.0 $\pm$ 3.5 & 9.1 & 0.3 	& 0.7 & -1.7 &   & $<$4.8 $\times 10^{-6}$ \\
185144 & 568.6 & 632.7 & 0.9 & 69.9 $\pm$ 12.7 & 72.0 & 1.0 	& 6.0 & -0.2 &   & $<$6.1 $\times 10^{-6}$ \\
186408 & 113.8 & 110.0 & 1.03 & 10.5 $\pm$ 5.7 & 14.0 & 0.8 	& 1.9 & -0.6 &   & $<$9.8 $\times 10^{-6}$ \\
186427 & 89.1 & 103.1 & 0.86 & -0.2 $\pm$ 5.6 & 11.6 & 0.0 	& 0.0 & -2.1 & & $<$4.3$\times 10^{-6}$ \\
188376 & 517.4 & 518.0 & 1.00 & 44.9 $\pm$ 13.3 & 58.5 & 0.8 	& 3.4 & -1.0 &   & $<$4.5 $\times 10^{-6}$ \\
189567 & 111.3 & 116.8 & 0.95 & 19.2 $\pm$ 3.3 & 13.3 & 1.4 	& 6.4 & 1.8 &   & $<$1.2 $\times 10^{-5}$ \\
190248 & 1270 & 1202 & 1.06 & 130.3 $\pm$ 21.2 & 133.9 & 1.0 	& 6.4 & -0.2 &   & $<$4.7 $\times 10^{-6}$ \\
196378 & 230.4 & 253.5 & 0.91 & 30.1 $\pm$ 5.6 & 28.9 & 1.0 	& 5.7 & 0.2 &   & $<$4.6 $\times 10^{-6}$ \\
197692 & 413.1 & 384.7 & 1.07 & 42.9 $\pm$ 9.2 & 43.5 & 1.0 	& 4.5 & -0.1 &   & $<$3.9 $\times 10^{-6}$ \\
203608 & 499.6 & 507.4 & 0.98 & 47.4 $\pm$ 9.7 & 57.8 & 0.8 	& 5.0 & -1.1 &   & $<$2.4 $\times 10^{-6}$ \\
206860$^{d}$ & 111.0 &115.8&0.96& 27.7 $\pm$ 3.8 & 13.1 & 2.1	& 8.1 & 3.9 & 16.8 & 1.1 $\times 10^{-5}$ \\
210277 & 83.5 & 91.9 & 0.91 & 8.0 $\pm$ 2.9 & 10.4 & 0.8 	& 3.1 & -0.8 & & $<$5.1$\times 10^{-6}$ \\
216437 & 107.5 & 102.1 & 1.05 & 9.5 $\pm$ 3.9 & 11.5 & 0.8 	& 2.8 & -0.5 & & $<$8.4$\times 10^{-6}$ \\
221420 & 135.2 & 125.3 & 1.08 & 15.6 $\pm$ 3.9 & 14.1 & 1.1 	& 4.5 & 0.4 &   & $<$9.4 $\times 10^{-6}$ 
\enddata 
\tablenotetext{a}{significance of excess (Eq.~\ref{chi70eq})}
\tablenotetext{b}{\70um dust fluxes have been color corrected by
15\%, appropriate for $\sim$50~K emission.}
\tablenotetext{c}{minimum $\ld$ from \70um emission (Eq.~\ref{ldeq})}
\tablenotetext{d}{star with excess \70um emission}
\tablenotetext{e}{star with excess \24um emission}
\label{resultstable}
\end{deluxetable} 

\begin{deluxetable}{lcccl}
\tablecaption{Selection bias: IRAS sources missing from our target list 
\label{legacytable}}  
\tablehead{Name & Spectral type & Distance (pc) & $\ld$ & References}
\startdata
$\epsilon$ Eri & K2V & 3.2 & $ 2.9 \times 10^{-4}$ & \hspace{0.04in} A, DD \\ 
$\tau$ Ceti & G8V & 3.6 & $2.5 \times 10^{-5}$ & \hspace{0.04in} Ha, DD\\ 
HD 17206 ($\tau^1$ Eri) & F6V  & 14.0 & $ 3.5 \times 10^{-4}$ &
\hspace{0.04in} A \\ 
HD 10647 & F8V & 17.3 & $5.4 \times 10^{-4}$ & \hspace{0.04in} SB, DD \\ 
\enddata 
\tablecomments{These sources meet our sample selection criteria
(\S\ref{sample}), but were observed by other guaranteed time programs.}
\tablerefs{See Table~\ref{reftable}}
\end{deluxetable}

\begin{deluxetable}{llc} 
\tablecaption{Reference abbreviations \label{reftable}}
\tablehead{ Symbol & Author & Values used}
\startdata
A     & \citet{aumann85} 			& IRAS phot \\ 
B    & \citet{barbieri2002} 			& age, [Fe/H] \\ 
Ba   & \citet{barry1988} 			& age \\ 
Bo   & \citet{borges1995}           		& [Fe/H] \\ 
C    & \citet{chen2001}                         & age, [Fe/H] \\ 
Ca   & \citet{destrobel1992,destrobel1997,destrobel2001} & age, [Fe/H] \\ 
Ce   & \citet{cenarro2001} 			& [Fe/H] \\ 
DD   & \citet{decin03}  			& ISO phot \\ 
E    & \citet{eggen1998} 			& [Fe/H] \\ 
F    & \citet{feltzing1998,feltzing2001} 	& age, [Fe/H] \\ 
Ga   & \citet{gaidos2002} 			& [Fe/H] \\ 
Ge   & \citet{degeus1990} 			& vis phot \\ 
Gi   & \citet{gimenez2000} 			& [Fe/H] \\ 
Go   & \citet{gonzalez2001} 			& [Fe/H] \\ 
Ha   & \citet{Habing01} 			& ISO phot \\
Hy   & \citet{haywood2001} 			& [Fe/H] \\ 
L    & \citet{lachaume1999}	 		& age, [Fe/H] \\ 
Le   & \citet{lebreton1999} 			& [Fe/H] \\ 
Lw   & \citet{laws2003}				& age, [Fe/H] \\ 
M    & \citet{marsakov1988,marsakov1995}	& age, [Fe/H] \\ 
Ma   & \citet{malagnini2000} 			& [Fe/H] \\ 
Ms   & \citet{mashonkina2001} 			& [Fe/H] \\ 
Mu   & \citet{munari1999} 			& [Fe/H] \\ 
N    & \citet{nordstrom04} 			& age,[Fe/H] \\ 
R    & \citet{randich1999} 			& age, [Fe/H] \\ 
Sa   & \citet{santos01} 			& [Fe/H] \\ 
SB   & \citet{stencel91} 			& IRAS phot \\ 
So   & \citet{song2000} 			& age \\ 
T    & \citet{taylor1994} 			& [Fe/H] \\ 
W    & \citet{Wichmann2003}			& age \\ 
Wr   & \citet{Wright04} 	 		& age 
\enddata 
\end{deluxetable} 

\end{document}